\documentclass[12pt]{article} %\input epsf.tex
\usepackage{graphicx,subfigure,latexsym}
\usepackage{amsmath,amssymb,amsfonts}
\usepackage{bm}

\newcommand{\be}{\begin{equation}}
\newcommand{\ee}{\end{equation}}
\newcommand{\bear}{\begin{eqnarray}}
\newcommand{\eear}{\end{eqnarray}}
\newcommand{\ba}{\begin{array}}
\newcommand{\ea}{\end{array}}

\def \be {\begin{equation}}
\def \ee {\end{equation}}

\def \bes {\begin{subequations}}
\def \ees {\end{subequations}}

\def \<{\langle}
\def \>{\rangle}
\def \+{\dagger}
\def \({\left(}
\def \){\right)}
\def \[{\left[}
\def \]{\right]}

\def \pl {\parallel}

\begin{document}

\begin{titlepage}
\vfill
\begin{flushright}
{\normalsize RBRC-1158}\\
\end{flushright}

\vfill
\begin{center}
{\Large\bf  Spin Polarized Photons from Axially Charged Plasma at Weak Coupling: Complete Leading Order }

\vskip 0.3in

\vskip 0.3in
Kiminad A. Mamo$^{1}$\footnote{e-mail: {\tt kabebe2@uic.edu}} and
Ho-Ung Yee$^{1,2}$\footnote{e-mail: {\tt hyee@uic.edu}}
\vskip 0.15in

{\it $^{1}$ Department of Physics, University of Illinois, Chicago, Illinois 60607}\\[0.15in]
{\it $^{2}$ RIKEN-BNL Research Center, Brookhaven National Laboratory,}\\
{\it Upton, New York
11973-5000}\\[0.15in]
{\normalsize  2015}

\end{center}

\vfill

\begin{abstract}

In the presence of (approximately conserved) axial charge in the QCD plasma at finite temperature, the emitted photons are spin-aligned, which is a unique P- and CP-odd signature of axial charge in the photon emission observables. We compute this ``P-odd photon emission rate" in weak coupling regime at high temperature limit  to complete leading order in the QCD coupling constant: the leading log as well as the constant under the log.
As in the P-even total emission rate in the literature, the computation of P-odd emission rate at leading order consists of three parts: 1) Compton and Pair Annihilation processes with hard momentum exchange, 2) soft t- and u-channel contributions with Hard Thermal Loop re-summation, 3) Landau-Pomeranchuk-Migdal (LPM) re-summation of collinear Bremstrahlung and Pair Annihilation. We present analytical and numerical evaluations
of these contributions to our P-odd photon emission rate observable.

\end{abstract}

\vfill

\end{titlepage}
\setcounter{footnote}{0}

\baselineskip 18pt \pagebreak
\renewcommand{\thepage}{\arabic{page}}
%\tableofcontents
\pagebreak

\section{Introduction}

Possible fluctuation of axial charge in QCD plasma through topological color field configurations, either from initial color
glass fields \cite{Kharzeev:2001ev} or from thermal sphaleron transitions, is one of the fundamental aspects of QCD dynamics.
Axial charge is both P- and CP-odd, and this distinct symmetry entails several interesting and unique phenomena
associated to it, such as Chiral Magnetic Effect \cite{Kharzeev:2007tn,Kharzeev:2007jp,Fukushima:2008xe}. In Ref.\cite{Mamo:2013jda} we explored and classified possible P- and CP-odd observables in photon and di-lepton emission rates, and found that the P- and CP-odd signals can be encoded in spin asymmetries of emitted photons and di-leptons.
See Ref.\cite{Ipp:2007ng} for a study of spin polarization of photons from a rotating plasma. Denoting the photon emission rate with fixed photon helicity $h=\pm 1$ (that is, spin alignment along the momentum) as $\Gamma^\pm$, the unique P- and CP-odd photon observable is\footnote{The $\Gamma^\pm$ can be the differential rates in momentum space.}
\be
A_{\pm\gamma}\equiv {{\Gamma^+}-{\Gamma^-}\over {\Gamma^+}+{\Gamma^-}}\,.\label{photon}
\ee

For di-leptons, let $\Gamma^{s_1,s_2}$ be the rate with fixed helicities $(s_1,s_2)=\left(\pm{1\over 2},\pm{1\over 2}\right)$ of a lepton and anti-lepton pair respectively, and the P- and CP-odd observable is given by
\be
A_{\pm l\bar l}\equiv {\Gamma^{+{1\over 2},+{1\over 2}}-\Gamma^{-{1\over 2},-{1\over 2}}\over
\Gamma^{+{1\over 2},+{1\over 2}}+\Gamma^{-{1\over 2},-{1\over 2}}}\,,\label{dilepton}
\ee
whereas the total di-lepton emission rate (at a given momentum bin) is
\be
\Gamma_{l\bar l}=\Gamma^{+{1\over 2},+{1\over 2}}+\Gamma^{+{1\over 2},-{1\over 2}}+\Gamma^{-{1\over 2},+{1\over 2}}+\Gamma^{-{1\over 2},-{1\over 2}}\,.
\ee

As these observables share the same P- and CP-odd parities with the axial charge, their signals naturally arise from the axial charge of the QCD plasma. QCD is a P- and CP-even theory, and the axial charge can only exist as
temporal and local fluctuations. The relaxation rate of axial charge via sphaleron transitions in a deconfined QCD plasma at weak coupling is
given by \cite{Bodeker:1998hm,Arnold:1998cy}
\be
\tau_R^{-1}={(2N_F)^2 \Gamma_{\rm sph}\over 2\chi T}\sim \alpha_s^5\log(1/\alpha_s) T\,,
\ee
where $\Gamma_{\rm sph}$ is the sphaleron rate and $\chi$ is the charge susceptibility. The effect of small quark mass $m_q$ to the relaxation rate is expected to be $\sim \alpha_s m_q^2/T$ \cite{unpub}.
On the other hand, the photon and di-lepton emission rates for hard momenta comparable to $T$ are $d\Gamma/d^3 k\sim \alpha_{\rm EM}\alpha_s\log(1/\alpha_s) T$ at leading order. We will assume in our work that $\alpha_s^5, \alpha_s (m_q/T)^2 \ll \alpha_{\rm EM}\alpha_s$ at sufficiently high temperature, so that the axial charge, once created by initial conditions or fluctuations, stays long enough to justify our computation of the above P- and CP-odd observables at weak coupling in the presence of an approximately constant value of axial chemical potential in the massless chiral limit.
In this work, we will present the computation of $A_{\pm\gamma}$ for photons with hard momenta at complete leading order in $\alpha_s$, and postpone a computation of di-lepton observable $A_{\pm l\bar l}$ to a future study.

In heavy-ion experiments, since the axial charge fluctuation averages to zero over many events, our observables should be measured either on the event-by-event basis, or one can look at the average of the squared. If the latter is chosen, one needs to take care of possible background fluctuations as well.

In Ref.\cite{Mamo:2013jda}, we derived explicit expressions relating the axial chemical potential to
our P- and CP-odd observables (\ref{photon}), (\ref{dilepton}).
Letting the momentum direction of a photon be along $\hat x^3$, and defining $G^R_\pm\equiv (G^R_{11}\pm i G^R_{12})$ (rotational invariance dictates that $G^R_{11}=G^R_{22}$ and $G^R_{12}=-G^R_{21}$) where $G^R_{ij}$ is the retarded correlation function of electromagnetic current in momentum space\footnote{Our definition of currents does not include an explicit factor of $e$ in front, that is, they are ``number'' currents.}
\be
G^R_{ij}(k)=(-i)\int d^4 x\,\,e^{-ikx}\theta(x^0)\langle [J_i(x),J_j(0)]\rangle\,,
\ee
we found
\be
{d\Gamma^\pm\over d^3 \bm k}={e^2\over (2\pi)^3 2\omega}n_B(\omega)(-2){\rm Im}\left[(\epsilon^\mu_\pm)^*\epsilon^\nu_\pm G^R_{\mu\nu}\right]={e^2\over (2\pi)^3 2\omega}n_B(\omega)(-2)
{\rm Im} G^R_\pm\,,\label{rate1}
\ee
for the emission rates with spin aligned polarization vectors
\be\epsilon^\mu_\pm={1\over\sqrt{2}}(0,1,\pm i,0)\,.\label{epsilon}\ee
Note that their sum is simply the total photon emission rate that has been computed in literature.
The difference that appears in our observable $A_{\pm\gamma}$ is given by
\be
{d\Gamma^{\rm odd}\over d^3\bm k}\equiv {d\Gamma^+\over d^3\bm k}-{d\Gamma^-\over d^3\bm k} = {e^2\over (2\pi)^3 2\omega}n_B(\omega)(-4)
{\rm Re} G^R_{12}\,.\label{poddemission}
\ee
We will refer $d\Gamma^{\rm odd}/ d^3\bm k$ simply as ``P-odd photon emission rate'' in the following.

The object $G^R_{12}(k)$ when $\bm k=|\bm k|\hat x^3$ arises from the P-odd part of the retarded correlation functions.
Rotational invariance and Ward identity allow us to have a unique P-odd structure in addition to the usual P-even part,
\be
G^R_{ij}(k)\sim i\sigma_\chi(k)\epsilon^{ijl}\bm k^l\,,\label{cmc}
\ee
which is in fact responsible for the Chiral Magnetic Effect at finite frequency-momentum $k$ of the external magnetic field \cite{Kharzeev:2009pj,Yee:2009vw},
\be
\bm J=\sigma_\chi(k)e\bm B(k)\,.
\ee
Since ${\rm Re} G^R_{12}(k)=-{\rm Im} \sigma_\chi(k)$, the P-odd emission rate $d\Gamma^{\rm odd}/d^3\bm k$ measures the imaginary part of chiral magnetic conductivity $\sigma_\chi(k)$ at light-like momenta. For small values of axial chemical potential, the chiral magnetic conductivity, and hence the P-odd emission rate, is proportional to the axial chemical potential. In our present study, although our results and expressions are in full dependency on axial chemical potential beyond linear order, we will present our numerical results only for linear dependency.

Note that the Chiral Magnetic Effect at zero momentum limit that has been shown to be universal,
\be
\lim_{k\to 0} \sigma_\chi(k)={N_c\over 2\pi^2}\left(\sum_F Q_F^2\right)\mu_A\equiv \sigma_0\,,\label{cme}
\ee
{\it does not} contribute to the imaginary part of $\sigma_\chi(k)$, and the P-odd photon emission rate is insensitive to this topological result. The imaginary part of $\sigma_\chi(k)$ is a dynamics driven quantity, and is highly sensitive to microscopic content and interactions of the theory.
For example, its small frequency limit at zero spatial momentum was recently computed in Ref.\cite{Jimenez-Alba:2015bia} at leading log order in the QCD coupling $\alpha_s={g^2/(4\pi)}$ to find
\be
{\rm Im}\sigma_\chi(\omega,\bm 0)=-\xi^{\rm QCD}_5\omega+{\cal O}(\omega^3)\,,\quad \xi^{\rm QCD}_5=-{2.003\over g^4\log(1/g)}{\mu_A\over T}\,,
\ee
which appears in the first time-derivative correction to the Chiral Magnetic Effect as
\be
\bm J=\sigma_0 e\bm B+\xi^{\rm QCD}_5 e{d\bm B\over dt}+\cdots\,.\label{timecme}
\ee
The computation of $\xi_5$ shares many common features with that of the ordinary electric conductivity (which also has  $\sim 1/(g^4\log(1/g))$ behavior), and is sensitive to the same QCD dynamics that the electric conductivity is subject to.
Nonetheless it relies on the existence of axial chemical potential, dictated by P- and CP-odd parities.

In the next section \ref{sec2}, we will formalize the dynamical nature of ${\rm Im}\sigma_\chi(k)$ by introducing the concept of ``P-odd spectral density'', which naturally appears in the fluctuation-dissipation relation of P-odd part of current correlation functions.
The section \ref{sec3} presents the main steps and results of our computation of the P-odd photon emission rate $d\Gamma^{\rm odd}/d^3\bm k$ at complete leading order in $\alpha_s$. We summarize and discuss our results in section \ref{sec4}.

\section{P-odd Spectral Density \label{sec2}}

One can formalize the dynamical nature of the imaginary part of chiral magnetic conductivity by the concept of ``P-odd spectral density'', first introduced in Ref.\cite{Jimenez-Alba:2015bia} (see Appendix 1 of that reference).
We choose to discuss it in real-time Schwinger-Keldysh formalism, where we have two time contours joined at future infinity, one is going forward in time (labeled as contour 1) and the other is going backward (contour 2). Initial thermal
density matrix is realized by attaching an imaginary time thermal contour at the beginning time (at past infinity).
By placing operators in suitable positions in the two contours, one can generate all kinds of time orderings for correlation functions. In terms of ``ra''-variables defined by
\be
{\cal O}_r={1\over 2}\left({\cal O}_1+{\cal O}_2\right)\,,\quad {\cal O}_a={\cal O}_1-{\cal O}_2\,,
\ee
our starting point is the thermal relation for the current-current correlation functions
\be
G^{rr}_{ij}(k)=\left({1\over 2}+n_B(k^0)\right)\left(G^{ra}_{ij}(k)-G^{ar}_{ij}(k)\right)\,.\label{thermal}
\ee
The retarded Green's function is given in this notation by
\be
G^R_{ij}(k)=-i G^{ra}_{ij}(k)\,,
\ee
and by hermiticity of the current operator, the retarded Green's function should be real-valued in coordinate space.
This requires to have $(G^R_{ij}(k))^*=G^R_{ij}(-k)$ in momentum space, or equivalently
\be
(G^{ra}_{ij}(k))^*=-G^{ra}_{ij}(-k)\,.\label{reality}
\ee
On the other hand, by definition, $G^{ra}_{ij}(x)=G^{ar}_{ji}(-x)$, so that in momentum space we have
\be
G^{ar}_{ij}(k)=G^{ra}_{ji}(-k)=-(G^{ra}_{ji}(k))^*\,,\label{reality2}
\ee
where the last equality comes from (\ref{reality}).

In the relation (\ref{thermal}), the left-hand side means the fluctuation amplitude, and the right-hand side, besides the statistical factor, represents the spectral density
\be
G^{rr}_{ij}(k)=\left({1\over 2}+n_B(k^0)\right)\rho_{ij}(k)\,,\quad \rho_{ij}(k)\equiv G^{ra}_{ij}(k)-G^{ar}_{ij}(k)\,.
\ee
The relation (\ref{reality2}) gives us
\be
\rho_{ij}(k)=G_{ij}^{ra}(k)+(G^{ra}_{ji}(k))^*\,,\label{herm}
\ee
so that the spectral density is twice of the hermitian part of $G^{ra}_{ij}(k)$ in terms of spatial $i,j$ indices. In a P-even ensemble, rotational invariance dictates that $G^{ra}_{ij}(k)$ be proportional to $\delta^{ij}$ or $\bm k^i \bm k^j$, and hence be symmetric with respect to $i,j$. The resulting spectral density from this should then be real-valued by (\ref{herm}).

In a P-odd ensemble, such as with axial chemical potential, rotational invariance allows us to have a purely imaginary and anti-symmetric (and hence hermitian) spectral density,
\be
\rho_{ij}(k)\sim \rho^{\rm odd}(k) i\epsilon^{ijl}\bm k^l\,,
\ee
with a real valued function $\rho^{\rm odd}(k)$. From (\ref{cmc}), we have $\rho^{\rm odd}(k)=-2 {\rm Im} \sigma_\chi(k)$, that is, the P-odd spectral density is in fact the imaginary part of chiral magnetic conductivity. We see that the imaginary part of chiral magnetic conductivity governs P-odd thermal fluctuations of currents, while the topological real part at zero momentum limit (\ref{cme}) does not contribute to thermal fluctuations. This gives some intuition why ${\rm Im}\sigma_\chi(k)$ is subject to microscopic real-time dynamics of the theory.

From (\ref{reality}), and (\ref{herm}), we have
\be
\rho^{\rm odd}(-k)=-\rho^{\rm odd}(k)\,.
\ee
Rotational invariance dictates that $\rho^{\rm odd}(k)$ be a function of $|\bm k|$, so $\rho^{\rm odd}(\omega,|\bm k|)$
is an odd function on $\omega$, similarly to P-even spectral densities. In small frequency, zero momentum limit we expect to have
\be
\rho^{\rm odd}(\omega, \bm 0)\sim 2\xi_5\omega+\cdots\,,\quad\omega\to 0\,,
\ee
where the hydrodynamic transport coefficient $\xi_5$ has the meaning of (\ref{timecme}). As the sign of $\xi_5$ depends
both on the chirality and the axial chemical potential, there seems to be no concept of positivity constraint on it, contrary to electric conductivity. However, explicit computations indicate that the ``relative'' sign between $\sigma_0$ (defined in (\ref{timecme})) and $\xi_5$ is always negative, reminiscent of magnetic induction \cite{Jimenez-Alba:2015bia}. We are not yet aware of any formal proof on this.

Our P-odd photon emission rate is related to the P-odd spectral density via (\ref{poddemission}) by
\be
{d\Gamma^{\rm odd}\over d^3\bm k}=-{e^2\over (2\pi)^3}n_B(\omega)\rho^{\rm odd}(\omega,\bm k)\big|_{\omega=|\bm k|}\,,
\ee
which explains that the P-odd photon emission rate, while it is P- and CP-odd, is a dynamics driven observable.

\section{P-odd Photon Emission Rate at Weak Coupling: Complete Leading Order\label{sec3}}

In this section, we compute the P-odd photon emission rate at complete leading order in QCD coupling $\alpha_s$,
\be
{d\Gamma^{\rm odd}\over d^3\bm k}\equiv {d\Gamma^+\over d^3\bm k} -{d\Gamma^-\over d^3\bm k}\sim \alpha_{EM}\alpha_s(\log(1/\alpha_s)+c)\,,
\ee
with an (approximate) axial chemical potential $\mu_A$ in the chiral limit of QCD.

A massless Dirac quark consists of a pair of left- and right-handed Weyl fermions. At leading order in $\alpha_s$,
the QCD interaction between them gives a higher order correction to the photon emission rate, and hence we can treat them independently. This will be clear in the Feynman diagrams we compute in the following. The only effect of having the other chiral Weyl fermion appears in the value of Debye mass $m_D^2$ in the gluon Hard Thermal Loop self-energy which enters the Landau-Pomeranchuk-Migdal (LPM) resummation of collinear Bremstrahlung and pair-annihilation that we compute in subsection \ref{subsec4}.
We therefore present our computational details only for the right-handed Weyl fermion with its chemical potential $\mu=\mu_A$.
The other left-handed Weyl fermion then has $\mu=-\mu_A$, and the total contribution to our P-odd photon emission rate is simply twice of that from the right-handed Weyl fermion, up to the above mentioned modification of $m_D^2$.
We assume our Dirac quark has a electromagnetic charge $Q=+1$, and the full result for two flavor QCD is simply
\be
Q_u^2+Q_d^2={5\over 9}\,,
\ee
times of the result for $Q=+1$ (where again $m_D^2$ has to include two flavor contributions).

We briefly summarize our notation and convention for a right-handed Weyl fermion theory.
Our metric convention is $\eta=(-,+,+,+)$. Let us define
\be
\sigma^\mu=({\bf 1},\bm\sigma)\,,\quad\bar\sigma^\mu=({\bf 1},-\bm\sigma)\,,
\ee
which satisfy
\be
\sigma^\mu\bar\sigma^\nu+\bar\sigma^\mu\sigma^\nu=-2\eta^{\mu\nu}\,.
\ee
The equation \be(p\cdot\sigma)(p\cdot\bar\sigma)=-p^2=(p^0)^2-|\bm p|^2\,,\ee
and the following trace formula will be useful,
\be
{\rm Tr}(\sigma^\mu\bar\sigma^\nu\sigma^{\alpha}\bar\sigma^\beta)=2(\eta^{\mu\nu}\eta^{\alpha\beta}+\eta^{\mu\beta}\eta^{\nu\alpha}-\eta^{\mu\alpha}\eta^{\nu\beta}+i\epsilon^{\mu\nu\alpha\beta})\,.\label{trace2}
\ee
The right-handed Weyl fermion action with QCD coupling $g$ is
\be
{\cal L}=i\psi^\dagger \sigma^\mu(\partial_\mu-ig t^a A_\mu^a)\psi\,,
\ee
Upon quantization, we have
\be
\psi(x)=\int{d^3\bm p\over \sqrt{2|\bm p|}}\left(u(\bm p) a_{\bm p} e^{-i|\bm p|t+i\bm p\cdot\bm x}+v(\bm p)b^\dagger_{-\bm p} e^{i|\bm p|t+i\bm p\cdot\bm x}\right)\,,
\ee
where particle and antiparticle spinors are defined by
\be
({\bf 1}-\bm\sigma\cdot\hat{\bm p})u(\bm p)=0\,,\quad ({\bf 1}+\bm \sigma\cdot\hat {\bm p})v(\bm p)=0\,,\quad \hat {\bm p}\equiv{\bm p\over |\bm p|}\,,
\ee
with normalization
\be
u(\bm p)u^\dagger(\bm p)=-p\cdot\bar\sigma\,,\quad v(\bm p)v^\dagger(\bm p)=-p\cdot\sigma\,,\quad p^\mu=(|\bm p|,\bm p)\,.\label{norm}
\ee
Note also that $ v(-\bm p)v^\dagger(-\bm p)=-p\cdot\bar\sigma$. It will be convenient to define spin projection operators to quark/anti-quark states
\be
{\cal P}_s(\bm p)\equiv{1\over 2}\left({\bf 1}+s \hat{\bm p}\cdot\bm\sigma\right)=-s{p_s\cdot\bar\sigma\over 2|\bm p|}\,,\quad p_s\equiv (s|\bm p|,\bm p)\,,\quad s=\pm 1\,,
\ee
in terms of which the (bare) real-time propagators in ``r/a'' basis are
\bear
S^{ra}(p)&=&i{p\cdot\bar\sigma\over p^2}\bigg|_{p^0\to p^0+i\epsilon}=\sum_{s=\pm}{i\over p^0-s|\bm p|+i\epsilon}{\cal P}_s(\bm p)\,,\nonumber\\
S^{ar}(p)&=&\sum_{s=\pm}{i\over p^0-s|\bm p|-i\epsilon}{\cal P}_s(\bm p)\,,\nonumber\\
S^{rr}(p)&=& \left({1\over 2}-n_+(p^0)\right)\left(S^{ra}(p)-S^{ar}(p)\right)=\left({1\over 2}-n_+(p^0)\right)\rho_F(p)\,,\label{barera}
\eear
where $n_\pm(p^0)=1/(e^{\beta(p^0\mp \mu)}+1)$ and the (bare) fermionic spectral density is
\be
\rho_F(p)=(2\pi)\sum_{s=\pm}\delta(p^0-s|\bm p|){\cal P}_s(\bm p)\,.\label{barera2}
\ee

The Feynman rules are as usual, for example,
for incoming (out-going) quark of momentum $\bm p$, we have $u(\bm p)$ ($u^\dagger(\bm p)$), and for the incoming (out-going) antiquark of momentum $\bm p$, we have $v^\dagger(-\bm p)$ ($v(-\bm p)$).
We remind ourselves of the rules for polarization states as it is important to get the correct sign for our P-odd photon emission rate. For out-going photon of polarization $\epsilon_\mu$, we attach $(\epsilon_\mu)^*$ contracted with the photon vertex $ie\sigma^\mu$ in the diagram. The same is true for gluons.
For incoming gluon of polarization $\tilde\epsilon_\mu$, we attach $\tilde\epsilon_\mu$ contracted with the gluon vertex $ig t^a\sigma^\mu$.
Finally, with these normalizations, the natural momentum integration measure is
\be
\int{d^3\bm p\over (2\pi)^3 2|\bm p|}\,.
\ee

\subsection{Hard Compton and Pair Annihilation Contributions \label{subsec1}}

The leading order rate consists of three distinct contributions: 1) Compton and Pair Annihilation with hard (that is, comparable to $T$) momentum exchanges, 2) Soft (that is, much less than $T$) t-channel exchange contribution with
IR divergence regulated by Hard Thermal Loop (HTL) re-summation of exchanged fermion line, and 3) collinear Bremstrahlung and pair-annihilation contributions induced by multiple scatterings with soft thermal gluons, referred to as Landau-Pomeranchuk-Migdal (LPM) effect. The leading log result in $\alpha_s$ is produced by 1) and 2), and the matching of the two logarithms from 1) and 2) to have the cut-off dependence removed is an important consistency check for the computation. We will see that this happens for our result.

Our methods of computation for the above three contributions closely follow the well-known ones in literature \cite{Baier:1991em,Kapusta:1991qp,Arnold:2001ba,Arnold:2001ms}, and we apply them to our case of P-odd emission rate, modulo a few subtleties. The complexity of numerical evaluation
is somewhat heavier than the P-even total emission rate.

  \begin{figure}[t]
 \centering
 \includegraphics[height=5.5cm]{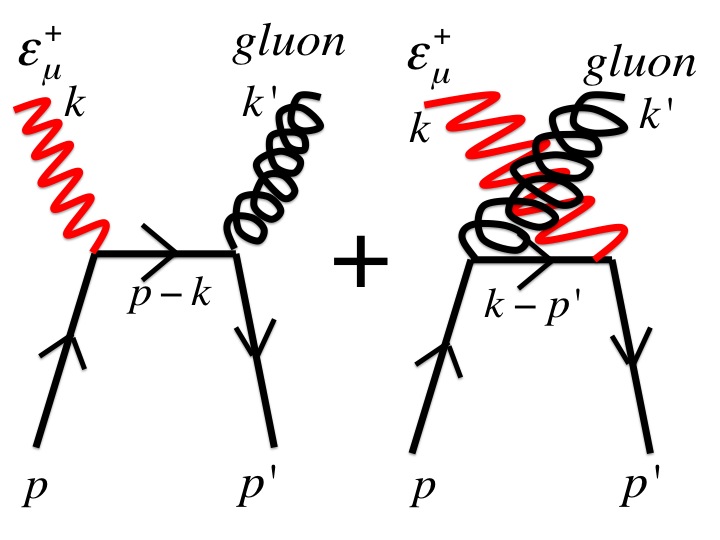}\caption{Pair Annihilation diagrams with hard momentum exchanges. \label{fig1}}
 \end{figure}
In this subsection, we describe hard Compton and Pair Annihilation rate computations.
Let the final photon momentum be $k$. For Pair Annihilation we label the momenta of incoming quark and antiquark pair by $p$ and $p'$ respectively, and let $k'$ be the momentum of out-going gluon of polarization $\tilde\epsilon^\mu$ and color $a$. There are two Feynman diagrams as in Figure \ref{fig1} with the total amplitude given as
\be
{\cal M}^{\rm pair}(\epsilon_\pm)=-ieg v^\dagger(-\bm p')\left[t^a \sigma^\nu{(p-k)\cdot\bar\sigma\over (p-k)^2}\sigma^\mu+\sigma^\mu {(k-p')\cdot\bar\sigma\over (k-p')^2}t^a\sigma^\nu\right]\,u(\bm p)(\epsilon^\pm_\mu)^*(\tilde\epsilon_\nu)^*\,,
\ee
where $\epsilon_\pm^\mu$ are the spin polarized photon states. Summing over colors in the squared amplitude produces a simple color factor
\be
\sum_a{\rm tr}(t^a t^a)=C_2(R) d_R={1\over 2}(N_c^2-1)\,,
\ee
for the fundamental representation of $SU(N_c)$. The summation over gluon polarization can be replaced by
\be
\sum_{\tilde\epsilon}(\tilde\epsilon_\nu)^*\tilde\epsilon_{\nu'}\to \eta_{\nu\nu'}\,,\label{ward}
\ee thanks to Ward identities. Since our P-odd photon emission rate is the difference between the rates with $\epsilon_+$ and $\epsilon_-$, what we need is the difference
\be
|{\cal M}^{\rm pair}(\epsilon_+)|^2-|{\cal M}^{\rm pair}(\epsilon_-)|^2\equiv |{\cal M}^{\rm pair}|^2_{\rm odd}\,,
\ee
and the Pair Annihilation contribution to the P-odd photon emission rate is written as
\bear
(2\pi)^3 2\omega {d\Gamma^{\rm odd}\over d^3\bm k}&=&\int {d^3\bm p\over (2\pi)^3 2|\bm p|}\int {d^3\bm p'\over (2\pi)^3 2|\bm p'|}\int {d^3\bm k'\over (2\pi)^3 2|\bm k'|} (2\pi)^4\delta(p+p'-k-k')\nonumber\\ &\times&  |{\cal M}^{\rm pair}|^2_{\rm odd}\,\,
n_+(|\bm p|) n_-(|\bm p'|)(1+n_B(|\bm k'|))\,.\label{pair}
\eear

The computation of P-odd amplitude $|{\cal M}^{\rm pair}|^2_{\rm odd}$ is algebraically complicated, although conceptually straightforward.
Using (\ref{norm}) and (\ref{ward}), and the polarization vectors
\be
\epsilon_\pm^\mu={1\over\sqrt{2}}(0,1,\pm i,0)\,,
\ee
after choosing $\bm k=|\bm k| \hat{\bm x}^3$, it reduces to computing traces of 8 $\sigma$ matrices. After some amount of efforts, we obtain a compact expression
\be
|{\cal M}^{\rm pair}|^2_{\rm odd}=C_2(R)d_R\cdot4e^2 g^2 (t-u)\left({1\over t}+{1\over u}-2\left({{\bm p}_\perp\over t}-{{\bm p}_\perp'\over u}\right)^2\right)\,,\label{pairamp}
\ee
where $t\equiv(p-k)^2$, $u\equiv(k-p')^2$, and ${\bm p}_\perp$ is the component of $\bm p$ perpendicular to the photon momentum $\bm k$.

The momentum integration in the emission rate (\ref{pair}) with the above P-odd amplitude possesses logarithmic IR divergences near $t\sim 0$ and $u\sim 0$,
corresponding to soft fermion exchanges. From the diagrams in Figure \ref{fig1}, it is clearly seen that the $u\sim 0$ divergence is the same type of divergence near $t\sim 0$ with a simple interchange of quark and anti-quark.
We can explore this symmetry of interchanging quark and anti-quark to simplify our computation:
the kinematics is identical under the interchange
\be
\bm p\longleftrightarrow \bm p'\,,\quad t\longleftrightarrow u\,,\quad n_+(|\bm p|)\longleftrightarrow n_-(|\bm p'|)\,,
\ee
and we can replace singular $\sim {1/ u}$ terms in the amplitude with $\sim {1/t}$ terms, so that the IR divergence appears in the new expression only around $t\sim 0$.
Explicitly, we can have a replacement
\bear
&& |{\cal M}^{\rm pair}|^2_{\rm odd}\,\,
n_+(|\bm p|) n_-(|\bm p'|)(1+n_B(|\bm k'|))\nonumber\\ &\longrightarrow&
C_2(R)d_R\cdot4e^2 g^2 \left(-{u\over t}-2(t-u)\left({{\bm p}_\perp^2\over t^2}-{{\bm p}_\perp\cdot{\bm p}_\perp'\over tu}\right)\right)\nonumber\\&\times&(n_+(|\bm p|) n_-(|\bm p'|)-n_-(|\bm p|)n_+(|\bm p'|))(1+n_B(|\bm k'|))\,.\label{pairA}
\eear
The integral with the above new expression has an additional advantage besides the absence of IR divergence near $u\sim 0$: from the new structure of distribution function factor, the fact that the result is an odd function on the chemical potential $\mu$ is manifest.

 \begin{figure}[t]
 \centering
 \includegraphics[height=5.5cm]{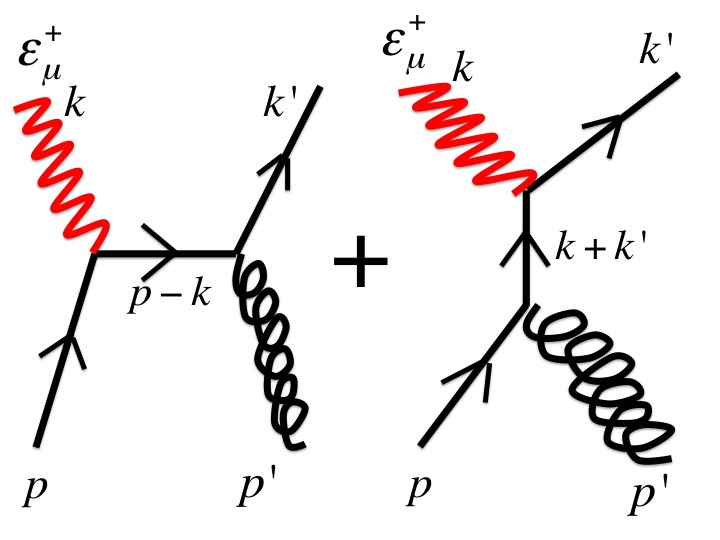}\caption{Compton scattering diagrams with hard momentum exchanges. \label{fig2}}
 \end{figure}
For Compton scatterings, let us first consider the Compton scattering with incoming quark of momentum $p$ and incoming gluon of momentum $p'$. The momentum of out-going quark will then be $k'$. The kinematics is identical to the Pair Annihilation case with the same definitions of $t\equiv (p-k)^2$, $u\equiv (k-p')^2$ and $s\equiv (k+k')^2$. Note that
\be
t+u+s=0\,.
\ee
There are two Feynman diagrams as in Figure \ref{fig2} with the amplitude
\be
{\cal M}^{\rm Compton}_{\rm quark}(\epsilon_\pm)=-iegu^\dagger(\bm k')\left[\sigma^\nu {(p-k)\cdot\bar\sigma\over (p-k)^2}\sigma^\mu+\sigma^\mu{(k+k')\cdot\bar\sigma\over (k+k')^2}\sigma^\nu\right]u(\bm p)(\epsilon_\mu^\pm)^*\tilde\epsilon_\nu\,,
\ee
where we omit color generators as it produces the same $C_2(R)d_R$ factor in the final result.
The P-odd amplitude square is then computed after some amount of algebra as
\bear
|{\cal M}^{\rm Compton}_{\rm quark}|^2_{\rm odd}&\equiv& |{\cal M}^{\rm Compton}_{\rm quark}(\epsilon_+)|^2-|{\cal M}^{\rm Compton}_{\rm quark}(\epsilon_-)|^2\nonumber\\
&=& C_2(R)d_R\cdot 4e^2 g^2 (s-t)\left({1\over t}+{1\over s}-2\left({{\bm p}_\perp\over t}+{{\bm k}_\perp'\over s}\right)^2\right)\,.\label{comptonamp}
\eear
The P-odd emission rate with this Compton amplitude for quarks is given by
\bear
(2\pi)^3 2\omega {d\Gamma^{\rm odd}\over d^3\bm k}&=&\int {d^3\bm p\over (2\pi)^3 2|\bm p|}\int {d^3\bm p'\over (2\pi)^3 2|\bm p'|}\int {d^3\bm k'\over (2\pi)^3 2|\bm k'|} (2\pi)^4\delta(p+p'-k-k')\nonumber\\ &\times&  |{\cal M}^{\rm Compton}_{\rm quark}|^2_{\rm odd}\,\,
n_+(|\bm p|)(1-n_+(|\bm k'|)) n_B(|\bm p'|)\,.\label{compton1}
\eear
There arises a logarithmic divergence near $t\sim 0$ only, which can be treated together with the one from the Pair Annihilation contribution.

The Compton scatterings with anti-quark has the P-odd amplitude square which is precisely negative to the above.
This could be expected simply from the fact that anti-quark has the opposite chirality (helicity) to that of quark, so P-odd observable has to flip sign between them. We confirmed this expectation by an explicit computation, but just for reference we present the Compton amplitude with anti-quark,
\be
{\cal M}^{\rm Compton}_{\rm antiquark}(\epsilon_\pm)=-ieg v^\dagger(-\bm p)\left[\sigma^\mu {(k-p)\cdot\bar\sigma\over (k-p)^2}\sigma^\nu+\sigma^\nu{(-k-k')\cdot\bar\sigma\over (k+k')^2}\sigma^\mu\right]v(-\bm k')(\epsilon_\mu^\pm)^*\tilde\epsilon_\nu\,.
\ee
Besides to this sign flip compared to the quark Compton contribution, the distribution function $n_+$ in (\ref{compton1}) has to be replaced by $n_-$ for anti-quarks, so the final Compton rate is given as
\bear
(2\pi)^3 2\omega {d\Gamma^{\rm odd}\over d^3\bm k}&=&\int {d^3\bm p\over (2\pi)^3 2|\bm p|}\int {d^3\bm p'\over (2\pi)^3 2|\bm p'|}\int {d^3\bm k'\over (2\pi)^3 2|\bm k'|} (2\pi)^4\delta(p+p'-k-k')\nonumber\\ &\times&  |{\cal M}^{\rm Compton}_{\rm quark}|^2_{\rm odd}\left(
n_+(|\bm p|)(1-n_+(|\bm k'|)) -n_-(|\bm p|)(1-n_-(|\bm k'|))\right)n_B(|\bm p'|)\,.\nonumber\\\label{compton2}
\eear
The fact the the result is an odd function on the chemical potential is also apparent here.

To perform the phase space integrations in (\ref{pair}) and (\ref{compton2}) with P-odd amplitudes (\ref{pairamp}) and (\ref{comptonamp}), we follow the technique nicely introduced and explained
in Refs.\cite{Baym:1990uj,Moore:2001fga}. The idea is to introduce auxiliary energy variable $q^0$ corresponding to either t-channel energy transfer (``t-channel parametrization'' according to Ref.\cite{Moore:2001fga}), or s-channel energy transfer (''s-channel parametrization'').
Its essential role is to trade the angular integration, coming from the energy $\delta$-function, for a scalar integration of $q^0$. The price to pay is a somewhat complicated, but manageable integration domain. The choice between t-channel and s-channel parametrizations is simply for convenience: t-channel parametrization is convenient for terms with $1/t$, and vice versa for s-parametrization.

We will give a brief summary on these parametrizations that one can also find in the original Refs.\cite{Baym:1990uj,Moore:2001fga}.
Let us focus on the common phase space integration measure in (\ref{pair}) and (\ref{compton2}),
\be
\int {d^3\bm p\over (2\pi)^3 2|\bm p|}\int {d^3\bm p'\over (2\pi)^3 2|\bm p'|}\int {d^3\bm k'\over (2\pi)^3 2|\bm k'|} (2\pi)^4\delta(p+p'-k-k')\,.
\ee
For t-channel parametrization, we perform $d^3\bm k'$ integration, and shift the integration variable $\bm p$ to $\bm q\equiv \bm p-\bm k$ to obtain\footnote{When we perform $\bm k'$ integration or shift the integration variable to $\bm q$, we should of course keep track of their effects in the amplitude and distribution function parts. We will present final summary on these parts as well.}
\be
\int{d^3\bm q\over (2\pi)^3 2|\bm q+\bm k| }\int{d^3\bm p'\over (2\pi)^3 2|\bm p'|}{1\over 2|\bm q+\bm p'|}(2\pi)\delta\left(|\bm q+\bm k|+|\bm p'|-|\bm k|-|\bm q+\bm p'|\right)\,.\label{phase1}
\ee
We then introduce a variable $q^0$ to write the energy $\delta$ function as
\be
\delta\left(|\bm q+\bm k|+|\bm p'|-|\bm k|-|\bm q+\bm p'|\right)=\int_{-\infty}^{+\infty}d q^0\,\,\delta\left(|\bm q+\bm k|-|\bm k|-q^0\right)\delta\left(q^0+|\bm p'|-|\bm q+\bm p'|\right)\,,
\ee
where the meaning of $Q\equiv (q^0,\bm q)$ as the t-channel exchange momentum is obvious.

\begin{figure}[t]
 \centering
 \includegraphics[height=5.9cm]{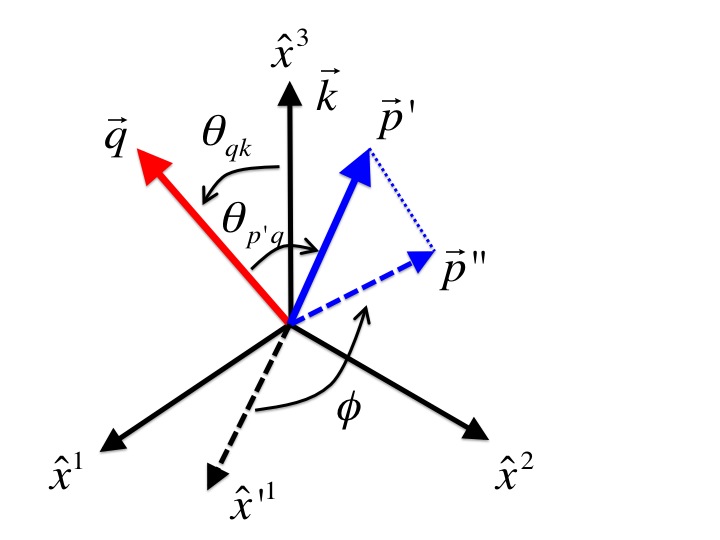}\caption{The geometry of t-channel parametrization. $(\hat{\bm q},\hat {\bm x}'^1,\hat {\bm x}^2)$ form an orthonormal basis rotated by $\theta_{\bm q\bm k}$, and ${\bm p}''$ is a projection of ${\bm p}'$ onto the $(\hat {\bm x}'^1,\hat {\bm x}^2)$ plane.  \label{fig3}}
 \end{figure}

\begin{figure}[t]
 \centering
 \includegraphics[height=5.9cm]{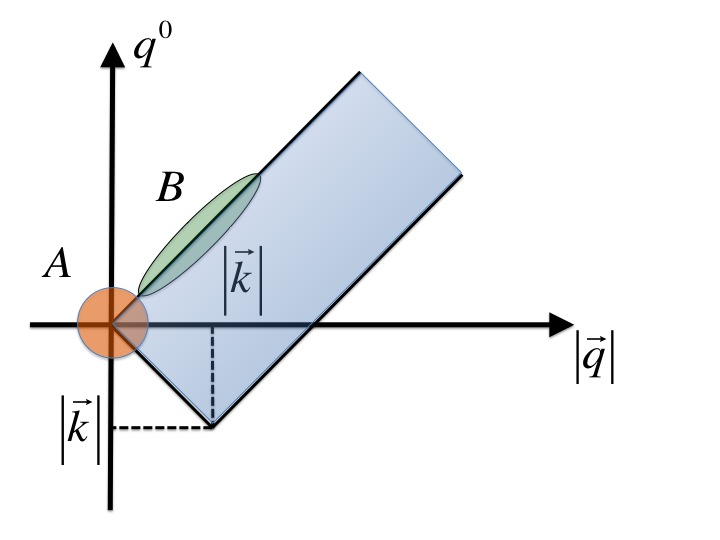}\caption{ The integration domain of $(q^0,|\bm q|)$ (shaded blue). The domain for $|\bm p'|$ is $|\bm p'|> (|\bm q|-q^0)/2$. The soft region $A$ (shaded red) is responsible for leading log IR divergence, and the region $B$ produces the energy logarithm that is described  in the following.\label{fig4}}
 \end{figure}

The next step is to express the energy $\delta$-functions in terms of angle variables. Denoting the angle between $\bm q$ and $\bm k$ as $\theta$, we have
\be
\delta\left(|\bm q+\bm k|-|\bm k|-q^0\right)={|\bm k|+q^0\over |\bm q||\bm k|}\delta\left(\cos\theta-\cos\theta_{\bm q\bm k}\right)\,,
\ee
where
\be
\cos\theta_{\bm q\bm k}={(q^0)^2-|\bm q|^2+2|\bm k|q^0\over 2|\bm q||\bm k|}\,.\label{qkangle}
\ee
There appears constraints on $(q^0,|\bm q|)$ simply from the requirement that $|\cos\theta_{\bm q\bm k}|\le 1$, which restricts the final integration domain that will be described shortly. Similarly, for the angle $\theta'$ between $\bm q$ and $\bm p'$ we have
\be
\delta\left(q^0+|\bm p'|-|\bm q+\bm p'|\right)={|\bm p'|+q^0\over |\bm p'||\bm q|}\delta\left(\cos\theta'-\cos\theta_{\bm p' \bm q}\right)\,,
\ee
with
\be
\cos\theta_{\bm p'\bm q}={(q^0)^2-|\bm q|^2+2|\bm p'|q^0\over 2 |\bm p'||\bm q|}\,.\label{pqangle}
\ee
Using these,
one can perform the angular integrals of $\cos\theta$ from $d^3\bm q$ and $\cos\theta'$ from $d^3 \bm p'$, localizing
$\cos\theta$ and $\cos\theta'$ to the values $\cos\theta_{\bm q\bm k}$ and $\cos\theta_{\bm p'\bm q}$.
Since we need to compute ${\bm p}_\perp={\bm q}_\perp$ and ${\bm p}_\perp'$ that appear in the P-odd amplitudes,
it is convenient to fix the photon momentum direction to be along $\hat{\bm x}^3$, and using the overall rotational symmetry in $(x^1,x^2)$-plane, we can align $\bm q$ to be in $(x^1,x^3)$ plane. See
Figure \ref{fig3} for the illustration.
 This alignment will produce a trivial $(2\pi)$ azimuthal integration factor in the integral of $d^3\bm q$.
Note that the azimuthal angle $\phi$ of $\bm p'$ with respect to $\bm q$ as defined in Figure \ref{fig3} still has to be integrated explicitly. From the geometry in Figure \ref{fig3}, we have
\be
{\bm q}_\perp=(|\bm q|\sin\theta_{\bm q\bm k},0)\,,\label{qperp}
\ee
in $(x^1,x^2)$ plane, and the ${\bm p}'$ in $(x^1,x^2,x^3)$-basis is given as
\bear
{\bm p}'&=&|\bm p'|\left(\begin{array}{ccc} \cos\theta_{\bm q\bm k}& 0 & \sin\theta_{\bm q\bm k}\\0&1& 0\\
-\sin\theta_{\bm q\bm k} & 0 & \cos\theta_{\bm q\bm k}\end{array}\right)\left(\begin{array}{c}\sin\theta_{\bm p'\bm q}\cos\phi\\\sin\theta_{\bm p'\bm q}\sin\phi\\\cos\theta_{\bm p'\bm q}\end{array}\right)\nonumber\\
&=&|\bm p'|\left(\begin{array}{c} \cos\theta_{\bm q\bm k}\sin\theta_{\bm p'\bm q}\cos\phi+\sin\theta_{\bm q\bm k}\cos\theta_{\bm p'\bm q}\\
\sin\theta_{\bm p'\bm q}\sin\phi\\
-\sin\theta_{\bm q\bm k}\sin\theta_{\bm p'\bm q}\cos\phi+\cos\theta_{\bm q\bm k}\cos\theta_{\bm p'\bm q}\end{array}\right)\,,\label{pperp}
\eear
which will be used in computing the P-odd amplitudes (\ref{pairamp}) and (\ref{comptonamp}). Finally,
the integration domain for $(q^0,|\bm q|,|\bm p'|)$ is depicted in Figure \ref{fig4}.

From all these, the phase space integration in the t-channel parametrization becomes
\bear
&&\int {d^3\bm p\over (2\pi)^3 2|\bm p|}\int {d^3\bm p'\over (2\pi)^3 2|\bm p'|}\int {d^3\bm k'\over (2\pi)^3 2|\bm k'|} (2\pi)^4\delta(p+p'-k-k')\nonumber\\
&=&{1\over  8 (2\pi)^4 |\bm k|}\int^\infty_0 d|\bm q|\int^{|\bm q|}_{{\rm max}(-|\bm q|,|\bm q|-2|\bm k|)}d q^0\int^{\infty}_{{|\bm q|-q^0\over 2}}d|\bm p'|\int^{2\pi}_0 d\phi\,.\label{tchannel}
\eear
For the amplitudes, we need to express various quantities in terms of integration variables and the angles $\theta_{\bm k\bm q}$ and $\theta_{\bm p'\bm q}$. The following expressions can be derived from (\ref{qperp}) and (\ref{pperp}) and the previous definitions:
\bear
&&t=-(q^0)^2+|\bm q|^2\,,\quad u=2|\bm k||\bm p'|\left(1+\sin\theta_{\bm q\bm k}\sin\theta_{\bm p'\bm q}\cos\phi-\cos\theta_{\bm q\bm k}\cos\theta_{\bm p'\bm q}\right)\,,\nonumber\\
&&{\bm q}_\perp^2=|\bm q|^2\sin^2\theta_{\bm q\bm k}\,,\quad {\bm q}_\perp\cdot{\bm p}_\perp'=|\bm q||\bm p'|\left(\sin\theta_{\bm q\bm k}\cos\theta_{\bm q\bm k}\sin\theta_{\bm p'\bm q}\cos\phi+\sin^2\theta_{\bm q\bm k}\cos\theta_{\bm p'\bm q}\right)\,,\nonumber\\
&&s=-t-u\,,\quad {\bm p}_\perp={\bm q}_\perp\,,\quad {\bm k}_\perp'={\bm q}_\perp+{\bm p}_\perp'\,,\label{tcq}
\eear
where $\theta_{\bm k\bm q}$ and $\theta_{\bm p'\bm q}$ are given by (\ref{qkangle}) and (\ref{pqangle}).
Finally, for the arguments that enter the distribution functions, we have
\be
|\bm p|=q^0+|\bm k|\,,\quad |\bm k'|=q^0+|\bm p'|\,.\label{tcq2}
\ee
The above data are enough, at least numerically, to compute the phase space integrations in (\ref{pair}) and (\ref{compton2}) to obtain our P-odd emission rate from the hard Compton and Pair Annihilation processes.
This t-channel parametrization is not efficient for the terms of $\sim 1/s$ or $\sim 1/s^2$ type, for which we use s-channel parametrization.

The geometry of s-channel parametrization is similar, so we simply summarize it.
The phase space measure becomes
\bear
&&\int {d^3\bm p\over (2\pi)^3 2|\bm p|}\int {d^3\bm p'\over (2\pi)^3 2|\bm p'|}\int {d^3\bm k'\over (2\pi)^3 2|\bm k'|} (2\pi)^4\delta(p+p'-k-k')\nonumber\\
&=&{1\over 8(2\pi)^4|\bm k|}\int^\infty_{|\bm k|} dq^0\int^{q^0}_{|2|\bm k|-q^0|}d|\bm q|
\int^{{q^0+|\bm q|\over 2}}_{{q^0-|\bm q|\over 2}} d|\bm p|\int^{2\pi}_0 d\phi\,,\label{schannel}
\eear
and we have
\bear
&&s=-(q^0)^2+|\bm q|^2\,,\quad t=2|\bm k||\bm p|\left(1+\sin\theta_{\bm q\bm k}\sin\theta_{\bm p\bm q}\cos\phi-\cos\theta_{\bm q\bm k}\cos\theta_{\bm p\bm q}\right)\,,\nonumber\\
&&{\bm q}_\perp^2=|\bm q|^2\sin^2\theta_{\bm q\bm k}\,,\quad {\bm p}_\perp\cdot{\bm q}_\perp=|\bm p||\bm q|\left(\sin\theta_{\bm q\bm k}\cos\theta_{\bm q\bm k}\sin\theta_{\bm p\bm q}\cos\phi+\sin^2\theta_{\bm q\bm k}\cos\theta_{\bm p\bm q}\right)\,,\nonumber\\
&& {\bm k'}_\perp={\bm q}_\perp\,,
\eear
where
\be
\cos\theta_{\bm q\bm k}={|\bm q|^2-(q^0)^2+2 q^0|\bm k|\over 2 |\bm q||\bm k|}\,,\quad\cos\theta_{\bm p\bm q}={|\bm q|^2-(q^0)^2+2 q^0|\bm p|\over 2|\bm q||\bm p|}\,,
\ee
and finally, we have to replace
\be
|\bm p'|\to q^0-|\bm p|\,,\quad |\bm k'|\to q^0-|\bm k|\,,
\ee
in the arguments of distribution functions.

The $\phi$ integrations in both t-channel and s-channel methods are at most of the type
\be
\int^{2\pi}_0 d\phi \,\,{A+B\cos\phi\over C+D\cos\phi}\,,
\ee
which can be done analytically.
The rest parts of the integration have to be done numerically, but we can identify the leading log parts of $\log(1/\alpha_s)$ and $\log(\omega/T)$ for $\omega\gg T$ analytically (recall $\omega=|\bm k|$), which we now describe.

\subsubsection{Leading Log}

The Pair Annihilation contribution (\ref{pair}) with (\ref{pairamp}) has a logarithmic IR divergence near $t\sim 0$, or
when $(q^0,|\bm q|)\ll |\bm k|, |\bm p'|$ in the t-channel parametrization. The same is true for the Compton rate (\ref{compton2}) with (\ref{comptonamp}).
These divergences are regulated by including HTL self-energy \cite{Braaten:1989mz} in the t-channel fermion propagator, which
screens the fermion exchange for soft momenta $(q^0,|\bm q|)\lesssim gT$ (``soft region'').
When $(q^0,|\bm q|)\gg gT$ (``hard region''), the HTL correction is sub-leading in $\alpha_s$ and what we have in the above as hard Compton and Pair Annihilation contributions give the leading order result.

A practical way to organize the leading order contributions from both regions is to introduce an intermediate scale
$gT\ll q^*\ll T$ \cite{Braaten:1991dd}, which serves as a t-channel IR cutoff for the above hard Compton and Pair Annihilation rates in the hard region, and as a t-channel UV cutoff for the same rates in the soft region with now the HTL self-energy included in the fermion propagator.
The latter soft region will be described in the next subsection \ref{subsec2}.
The two logs of $\log q^*$ from both regions have to match to produce a final result independent of $q^*$: after identifying $\log q^*$ from each region, we neglect $q^*/T$ and $(gT)/q^*$ corrections in the rest parts of the two regions by sending $q^*\to 0$ in the hard region and $q^*\to\infty$ in the soft region. The resulting (numerical) constant is the leading order constant under the log.

Let us identify the leading log from the hard region in this subsection. The t-channel parametrization is most efficient for this purpose. The $q^*$ is introduced as an IR cutoff of $d|\bm q|$-integral in (\ref{tchannel})\footnote{This meaning of $q^*$ has to be identical to the one in the soft region computation in subsection \ref{subsec2}.}:
\be
(2\pi)^3 2\omega {d\Gamma^{\rm odd}_{\rm hard}\over d^3\bm k}={1\over  8 (2\pi)^4 |\bm k|}\int^\infty_{q^*} d|\bm q|\int^{|\bm q|}_{{\rm max}(-|\bm q|,|\bm q|-2|\bm k|)}d q^0\int^{\infty}_{{|\bm q|-q^0\over 2}}d|\bm p'|\int^{2\pi}_0 d\phi\,\,{\cal I}\,,\label{finalhard10}
\ee
where ${\cal I}$ is the sum of the integrands in (\ref{pairA}) and (\ref{compton2}) from the Compton and Pair Annihilation processes:
\bear
{\cal I}&=&C_2(R)d_R\cdot4e^2 g^2 \left(-{u\over t}-2(t-u)\left({{\bm q}_\perp^2\over t^2}-{{\bm q}_\perp\cdot{\bm p}_\perp'\over tu}\right)\right)\nonumber\\&\times&(n_+(q^0+|\bm k|) n_-(|\bm p'|)-n_-(q^0+|\bm k|)n_+(|\bm p'|))(1+n_B(q^0+|\bm p'|))\nonumber\\ &+&C_2(R)d_R\cdot 4e^2 g^2 (s-t)\left({1\over t}+{1\over s}-2\left({{\bm q}_\perp\over t}+{({\bm q}_\perp+{\bm p}_\perp')\over s}\right)^2\right)\nonumber\\
&\times&\left(
n_+(q^0+|\bm k|)(1-n_+(q^0+|\bm p'|)) -n_-(q^0+|\bm k|)(1-n_-(q^0+|\bm p'|))\right)n_B(|\bm p'|)\,,\nonumber\\\label{A}
\eear
with the use of expressions in (\ref{tcq}) and (\ref{tcq2}) for the t-channel parametrization.

From the distribution functions, $|\bm p'|$ integral is dominated by $|\bm p'|\sim T$. The log divergence appears in small $(q^0,|\bm q|)\ll |\bm k|,|\bm p'|\sim T$ since we assume hard photons $T\lesssim |\bm k|$. Figure \ref{fig4} shows
this region (region $A$). In this case, from (\ref{qkangle}) and (\ref{pqangle}), we have
\be
\cos\theta_{\bm q\bm k}\approx \cos\theta_{\bm p'\bm q}\approx {q^0\over |\bm q|}\,,
\ee
and the leading behavior in ${\cal A}$ comes from the terms of $(u,s)/t$ or $(u,s){\bm q}_\perp^2/t^2$ types, which gives after some algebra,
\bear
{\cal I}&\sim& C_2(R)d_R\cdot 8 e^2 g^2 {|\bm k||\bm p'|\over |\bm q|^2}\left(1+\cos\phi\right) \nonumber\\
&\times&\left(n_+(|\bm k|)n_-(|\bm p'|)(1+n_B(|\bm p'|))+n_+(|\bm k|)n_B(|\bm p'|)(1-n_+(|\bm p'|) )- (n_+\leftrightarrow n_-)\right)\nonumber\\&=&C_2(R)d_R\cdot 8 e^2 g^2 {|\bm k||\bm p'|\over |\bm q|^2}\left(1+\cos\phi\right)\nonumber \\
&\times&\left(n_+(|\bm k|)n_-(0)-n_-(|\bm k|)n_+(0)\right)\left(n_+(|\bm p'|)+n_-(|\bm p'|)+2n_B(|\bm p'|)\right) \,,
\eear
where in the last line, we use an interesting identity
\be
n_\mp(|\bm p'|)(1+n_B(|\bm p'|))+n_B(|\bm p'|)(1-n_\pm(|\bm p'|))=n_\mp(0)\left(n_+(|\bm p'|)+n_-(|\bm p'|)+2n_B(|\bm p'|)\right)\,. \ee
We then have a leading log behavior
\bear
(2\pi)^3 2\omega {d\Gamma^{\rm odd}_{\rm hard}\over d^3\bm k}&\sim &
C_2(R)d_R\cdot{e^2 g^2\over (2\pi)^3}\left(n_+(|\bm k|)n_-(0)-n_-(|\bm k|)n_+(0)\right)\nonumber\\&\times&\int^{\sim T}_{q^*}d|\bm q|{1\over |\bm q|^2}
\int^{|\bm q|}_{-|\bm q|} dq^0\int_0^\infty d|\bm p'|\,|\bm p'|\left(n_+(|\bm p'|)+n_-(|\bm p'|)+2n_B(|\bm p'|)\right)\nonumber\\&\sim& C_2(R)d_R\cdot{e^2 g^2\over (2\pi)^3}\left(\pi^2 T^2+\mu^2\right)\left(n_+(|\bm k|)n_-(0)-n_-(|\bm k|)n_+(0)\right)\log\left({T/q^*}\right)\nonumber\\
&=& d_R {e^2\over (2\pi)}m_f^2\left(n_+(|\bm k|)n_-(0)-n_-(|\bm k|)n_+(0)\right)\log\left({T/q^*}\right)\,,\label{LLhard}
\eear
where we use
\be
\int_0^\infty d|\bm p'|\,|\bm p'|\left(n_+(|\bm p'|)+n_-(|\bm p'|)+2n_B(|\bm p'|)\right)={1\over 2}\left(\pi^2T^2+\mu^2\right)\,,
\ee and in the last line we write the result in terms of the asymptotic fermion thermal mass
\be
m_f^2=C_2(R) {g^2\over 4}\left(T^2+{\mu^2\over\pi^2}\right)\,.\label{fmass}
\ee
We will check that the leading log from the hard Compton and Pair Annihilation given in (\ref{LLhard}) nicely matches to the soft region result with HTL re-summation in the next subsection.

For an ultra-hard photon energy $\omega=|\bm k|\gg T$, there appears a logarithmic rise of $\log(\omega/T)$
in the energy dependence of the leading order constant under the log.
We close this subsection by identifying this ``energy logarithm''. For this aim, it is convenient to work with the light cone variables
\be
q^\pm\equiv {|\bm q|\pm q^0\over 2}\,,
\ee
with the measure change $d|\bm q| dq^0=2 dq^+ dq^-$. The energy logarithm appears in the domain where
\be
q^-\lesssim |\bm p'|\sim T \ll q^+ \ll |\bm k|=\omega\,,
\ee
which is also indicated in Figure \ref{fig4} (region $B$). In this case, we have
\be
\cos\theta_{\bm q\bm k}\approx {q^0\over |\bm q|}\approx 1\,,\quad \cos\theta_{\bm p'\bm q}={-4 q^+q^-+2q^0|\bm p'|\over 2|\bm p'||\bm q|}\approx 1-{2q^-\over |\bm p'|}\,,
\ee
and the leading behavior in $\cal A$ arises again from the same $(u,s)/t$ or $(u,s){\bm q}_\perp^2/t^2$ terms, with
\be
{\cal I}\sim C_2(R)d_R\cdot 4 e^2 g^2{|\bm k|\over q^+}\left(n_+(|\bm k|)\left(n_-(|\bm p'|)+n_B(|\bm p'|)\right)-(n_+\leftrightarrow n_-)\right)\,,
\ee
so that we have
\bear
&&(2\pi)^3 2\omega {d\Gamma^{\rm odd}_{\rm hard}\over d^3\bm k}\sim
C_2(R)d_R{e^2 g^2\over  (2\pi)^3}\int^{|\bm k|}_{\sim T}dq^+{1\over q^+}\int^\infty_0 dq^-\int_{q^-}^\infty d|\bm p'|\nonumber\\
&\times&\left(n_+(|\bm k|)\left(n_-(|\bm p'|)+n_B(|\bm p'|)\right)-(n_+\leftrightarrow n_-)\right)\nonumber\\
&=& C_2(R)d_R{e^2 g^2\over  (2\pi)^3}\log(|\bm k|/T)\left(n_+(|\bm k|)\int_0^\infty dq^-\,q^-\left(n_-(q^-)+n_B(q^-)\right)-(n_+\leftrightarrow n_-)\right)\,,\nonumber\\\label{energy}
\eear
where in the first line, we can safely let the upper cutoff of $q^-$ be infinity, due to the presence of effective cutoff by the distribution functions (more precisely, the cutoff is given by $\sim q^+\gg T$).

The integrals that appear in the above
\be
\int_0^\infty dq^-\,q^-\left(n_\mp(q^-)+n_B(q^-)\right)={T^2\over 6}\left(\pi^2-6\,\,{\rm Li}_2\left(-e^{\mp{\mu/T}}\right)\right)\,,
\ee
are not simple polynomials in $T$ and $\mu$, contrary to the case of leading log in coupling (\ref{LLhard}).

\subsection{Soft t-Channel Contribution: Hard Thermal Loop\label{subsec2} }

In this subsection, we compute the soft t-channel contributions from Compton and Pair Annihilation processes, whose IR divergence is regulated by re-summing fermion HTL self-energy in the fermion exchange line. Following the original treatment in Refs.\cite{Baier:1991em,Kapusta:1991qp}, we compute this directly in terms of 1-loop current-current correlation functions that enter the emission rate formula (\ref{rate1}) or (\ref{poddemission}), with one internal fermion line being soft, and hence HTL re-summed, corresponding to soft t-channel exchange. The emission rate written in (\ref{rate1}) is given by suitable imaginary part of the correlation functions, and by applying the cutting-rule, it is easy to see that the result should be equivalent to that from computing Feynman diagrams of only t-channel Compton and Pair Annihilation processes (with the HTL re-summed propagator) that we described in the previous subsection.

We compute the following with the soft t-channel momentum with an UV cutoff $q^*$,
\be
(2\pi)^3 2\omega{d\Gamma(\epsilon_\pm)\over d^3\bm k} = e^2 n_B(\omega)(-2)\,{\rm Im}\left[(\epsilon^\mu_\pm)^*\epsilon^\nu_\pm G^R_{\mu\nu}(k)\right]=e^2 n_B(\omega)\,2\,{\rm Re}\left[(\epsilon^\mu_\pm)^*\epsilon^\nu_\pm G^{ra}_{\mu\nu}(k)\right]\,.\label{poddrate1}
\ee
Since $(\epsilon^\mu_\pm)^*\epsilon^\nu_\pm$ is a hermitian matrix in terms of $\mu,\nu$ indices,
the emission rate picks up only the hermitian part of $G^{ra}_{\mu\nu}(k)$.
\begin{figure}[t]
 \centering
 \includegraphics[height=6cm]{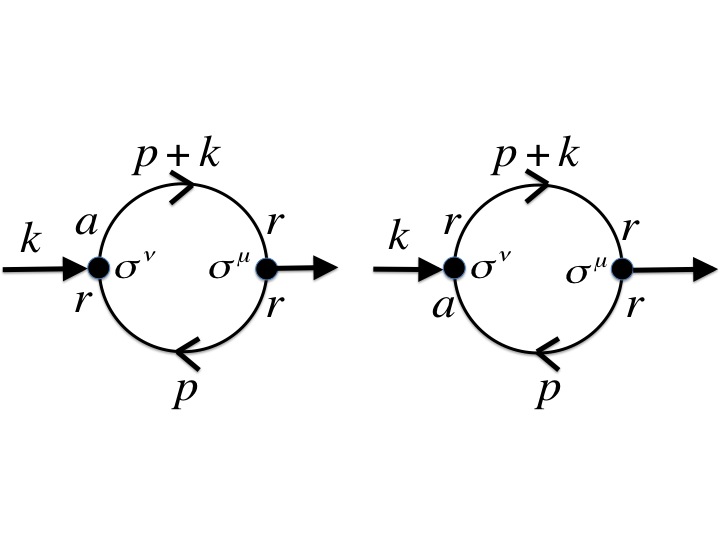}\caption{ Two real-time Feynman diagrams for $G^{ra}_{\mu\nu}(k)$ in the ``ra''-basis. \label{fig5}}
 \end{figure}
There are two real-time Feynman diagrams for $G^{ra}_{\mu\nu}(k)$ depicted in Figure \ref{fig5},
which gives
\be
G^{ra}_{\mu\nu}(k)=(-1)d_R\int{d^4 p\over (2\pi)^4}\,\,{\rm tr}\left[\sigma^\nu S^{rr}(p)\sigma^\mu S^{ra}(p+k)+\sigma^\nu S^{ar}(p)\sigma^\mu S^{rr}(p+k)\right]\,,
\ee
where $d_R$ is the dimension of color representation.
Recall the thermal relation
\be
S^{rr}(p)=\left({1\over 2}-n_+(p^0)\right)(S^{ra}(p)-S^{ar}(p))\equiv \left({1\over 2}-n_+(p^0)\right)\rho_F(p) \,,
\ee
and by the reality property $S^{ar}(p)^\dagger=-S^{ra}(p)$, $S^{rr}(p)$ and $\rho_F(p)$ are  hermitian matrices in terms of 2 component spinor indices. Using the same relations and the hermiticity of $\sigma^\mu$, it is easy to find the hermitian part of $G^{ra}_{\mu\nu}(k)$ as (we denote $\omega\equiv k^0=|\bm k|$)
\be
G^{ra}_{\mu\nu}(k)+(G^{ra}_{\nu\mu}(k))^*=d_R\int{d^4 p\over (2\pi)^4}\,\,\left(n_+(p^0)-n_+(p^0+\omega)\right){\rm tr}\left[\sigma^\nu\rho_F(p)\sigma^\mu\rho_F(p+k)\right]\,.\label{herG}
\ee
The emission rate is given solely by (fermion) spectral density $\rho_F$, which conforms to the expectation from cutting rules.

Bare fermion spectral density is easy to read off from (\ref{barera}) or (\ref{barera2}):
\be
\rho^{\rm bare}_F(p)=(2\pi)\sum_{s=\pm}\delta(p^0-s|\bm p|){\cal P}_s(\bm p)\,,
\ee
with the projection operators we repeat here for convenience,
\be
{\cal P}_s(\bm p)={1\over 2}\left({\bf 1}+s\hat{\bm p}\right)=-s{\bar\sigma\cdot p_s\over 2|\bm p|}\,,\quad p_s^\mu\equiv (s|\bm p|,\bm p)\,.
\ee
In general, fermion spectral density in a Weyl fermion theory including HTL  self-energy is written as (see Appendix 2 of Ref.\cite{Jimenez-Alba:2015bia}),
\be
\rho_F^{\rm HTL}(p)=\sum_s \rho^{\rm HTL}_s(p){\cal P}_s(\bm p)\,,\quad \rho^{\rm HTL}_s(p)=-2 \,{\rm Im}\left[{1\over p^0-s|\bm p|+\Sigma_s^{R, {\rm HTL}}(p)}\right]\,,
\ee
where the HTL self-energy is given by
\be
\Sigma_s^{R,{\rm HTL}}(p)=-{m_f^2\over 4|\bm p|}\left(2s+\left(1-s{p^0\over|\bm p|}\right)\log\left({p^0+|\bm p|+i\epsilon\over p^0-|\bm p|+i\epsilon}\right)\right)\,,
\ee
with the asymptotic fermion thermal mass that is introduced before in (\ref{fmass}),
\be
m_f^2=C_2(R){g^2\over 4}\left(T^2+{\mu^2\over\pi^2}\right)\,.
\ee

Inserting (\ref{herG}) into (\ref{poddrate1}), choosing the direction of $\bm k=|\bm k|\hat{\bm x}^3$ explicitly and computing the $\sigma$-matrix traces using (\ref{trace2}), we end up to an expression
for our P-odd emission rate as
\bear
(2\pi)^3 2\omega {d\Gamma^{\rm odd}\over d^3\bm k}&=&d_R e^2 n_B(\omega)\int {d^4p\over (2\pi)^4}\,\,
\left(n_+(p^0)-n_+(p^0+\omega)\right)\nonumber\\&\times&\sum_{s,t}\rho_s(p)\rho_t(p+k)\left(t{({\bm p}_3+|\bm k|)\over |\bm p+\bm k|}-s{{\bm p}_3\over |\bm p|}\right)\,,\label{form1}
\eear
where $\rho_{s,t}$ in the above can be either bare or HTL, depending on whether the momentum argument is hard or soft. We should consider the region of $p$ where one of the two momenta, $p$ or $p+k$, is soft, corresponding to soft t- or u-channel processes.

It would be convenient to combine the two soft regions into one, say soft $p$ region. That is, for soft $p+k$ region, let us change the variable $p\to -p-k$, so that in the new variable, $p$ is soft. Under this transform, we have
\be
n_+(p^0)-n_+(p^0+\omega)\to n_-(p^0)-n_-(p^0+\omega)\,,\quad \rho_s(p)\to\rho_{-s}(p+k)\,,\quad
\rho_t(p+k)\to \rho_{-t}(p)\,,
\ee
and relabeling $-t\to s$ and $-s\to t$, we arrive at the precisely the same form as in (\ref{form1}), with the replacement
\be
\left(n_+(p^0)-n_+(p^0+\omega)\right)\to -\left(n_-(p^0)-n_-(p^0+\omega)\right)\,,
\ee
therefore, we can study only the soft $p$ region of the following expression
\bear
(2\pi)^3 2\omega {d\Gamma^{\rm odd}_{\rm soft}\over d^3\bm k}&=&d_R e^2 n_B(\omega)\int {d^4p\over (2\pi)^4}\,\,
\left(n_+(p^0)-n_+(p^0+\omega)-(n_+\leftrightarrow n_-)\right)\nonumber\\&\times&\sum_{s,t}\rho^{\rm HTL}_s(p)\rho^{\rm bare}_t(p+k)\left(t{({\bm p}_3+|\bm k|)\over |\bm p+\bm k|}-s{{\bm p}_3\over |\bm p|}\right)\,,\label{form1}
\eear
where we explicitly indicated the HTL (bare) spectral density for soft (hard) $p$ ($p+k$).
An additional bonus is that the result is manifestly an odd function in the chemical potential. This is reminiscent of what happens in our previous computation of hard Compton and Pair Annihilation processes.

From
\be
\rho_t^{\rm bare}(p+k)=(2\pi)\delta(p^0+|\bm k|-t|\bm p+\bm k|)\,,
\ee
and since $p$ is soft while $(\omega=|\bm k|,\bm k)$ is hard, we see that only $t=1$ contributes.
The total integrand has a rotational symmetry on $(x^1,x^2)$-plane, so the azimuthal integral of $\bm p$ around $\bm k$ will trivially give $(2\pi)$. The polar integration can be done
by the same technique we use in (\ref{qkangle}): for $p\ll k$, we can write the integral measure including the energy $\delta$-function as
\be
\int {d^4 p\over (2\pi)^4} (2\pi)\delta(p^0+|\bm k|-|\bm p+\bm k|)={1\over (2\pi)^2}\int_0^\infty d|\bm p||\,|\bm p|\int_{-|\bm p|}^{|\bm p|} dp^0 \left(1+{p^0\over |\bm k|}\right)\bigg|_{{\bm p}_3\to |\bm p|\cos\theta_{\bm p\bm k}}\,,
\ee
where
\be
\cos\theta_{\bm p\bm k}={(p^0)^2-|\bm p|^2+2p^0|\bm k|\over 2|\bm p||\bm k|}\,.
\ee
Using this, our P-odd rate (\ref{form1}) from soft region is compactly written as
\bear
(2\pi)^3 2\omega {d\Gamma^{\rm odd}_{\rm soft}\over d^3\bm k}&=&d_R{e^2\over (2\pi)^2}n_B(\omega)\int_0^{q^*} d|\bm p||\,|\bm p|\int_{-|\bm p|}^{|\bm p|} dp^0 \left(1+{p^0\over |\bm k|}\right)\nonumber\\&\times&\left(n_+(p^0)-n_+(p^0+\omega)-(n_+\leftrightarrow n_-)\right)\nonumber\\&\times& \sum_s \rho_s^{\rm HTL}(p^0,|\bm p|)\left({|\bm p|\cos\theta_{\bm p \bm k}+|\bm k|\over p^0+|\bm k|}-s\cos\theta_{\bm p\bm k}\right)\,,
\eear
where we introduce the UV cutoff $q^*$ for the t-channel momentum integral of $|\bm p|$ to regulate the logarithmic diveregence. The meaning of $q^*$ here is identical to that used in the hard Compton and Pair Annihilation rates in the previous subsection, which is important to get the correct leading order constant under the log.

Since the cutoff is $q^*\ll T\lesssim |\bm k|$ (while $q^*\gg m_f\sim gT$), we have a further simplification at leading order to
\be
\cos\theta_{\bm p\bm k}\approx {p^0\over |\bm p|}\,,\quad \left({|\bm p|\cos\theta_{\bm p \bm k}+|\bm k|\over p^0+|\bm k|}-s\cos\theta_{\bm p\bm k}\right)\approx 1-s{p^0\over |\bm p|}\,,
\ee
and we arrive at
\bear
(2\pi)^3 2\omega {d\Gamma^{\rm odd}_{\rm soft}\over d^3\bm k}&\approx&d_R{e^2\over (2\pi)^2}n_B(\omega)\left(n_+(0)-n_+(\omega)-(n_+\leftrightarrow n_-)\right)\nonumber\\&\times&\int_0^{q^*} d|\bm p||\,|\bm p|\int_{-|\bm p|}^{|\bm p|} dp^0  \sum_s \rho_s^{\rm HTL}(p^0,|\bm p|)\left(1-s{p^0\over |\bm p|}\right)\nonumber\\
&=&d_R {e^2\over (2\pi)^2}\left(n_+(\omega)n_-(0)-n_-(\omega)n_+(0)\right)\nonumber\\&\times&\int_0^{q^*} d|\bm p||\,|\bm p|\int_{-|\bm p|}^{|\bm p|} dp^0  \sum_s \rho_s^{\rm HTL}(p^0,|\bm p|)\left(1-s{p^0\over |\bm p|}\right)\,,\label{softres}
\eear
where in the last line, we use an interesting identity
\be
n_B(\omega)(n_\pm(0)-n_\pm(\omega))=n_\pm(\omega)n_\mp(0)\,.
\ee

As it happens, the remaining integral is something that has been already computed in literature: the same integral appears in the P-even total emission rate.
In fact, a similar manipulation in our language produces the usual P-even total emission rate from soft t-channel region at leading order as
\bear
(2\pi)^3 2\omega {d\Gamma^{\rm total}_{\rm soft}\over d^3\bm k}&\approx&
d_R {e^2\over (2\pi)^2}\left(n_+(\omega)n_-(0)+n_-(\omega)n_+(0)\right)\nonumber\\&\times&\int_0^{q^*} d|\bm p||\,|\bm p|\int_{-|\bm p|}^{|\bm p|} dp^0  \sum_s \rho_s^{\rm HTL}(p^0,|\bm p|)\left(1-s{p^0\over |\bm p|}\right)\,,\label{pevenLO}
\eear
and matching to the known results in Refs.\cite{Baier:1991em,Arnold:2001ms} when $\mu=0$, we have at leading order
\be
\int_0^{q^*} d|\bm p||\,|\bm p|\int_{-|\bm p|}^{|\bm p|} dp^0  \sum_s \rho_s^{\rm HTL}(p^0,|\bm p|)\left(1-s{p^0\over |\bm p|}\right)
= (2\pi)m_f^2 \left(\log(q^*/m_f)-1+\log 2\right)\,.
\ee
Using this in  (\ref{softres}) we finally have the leading order expression for our P-odd emission rate as
\be
(2\pi)^3 2\omega {d\Gamma^{\rm odd}_{\rm soft}\over d^3\bm k}\approx
d_R {e^2\over (2\pi)}m_f^2\left(n_+(\omega)n_-(0)-n_-(\omega)n_+(0)\right)\left(\log(q^*/m_f)-1+\log 2\right)\,.\label{poddLO}
\ee

Nonetheless, it is instructive to see how the leading log arises from the above integral, using the sum rules for the fermion spectral densities $\rho_s^{\rm HTL}$.
The leading log comes from the region $m_f\ll |\bm p|\ll q^*$, and in this case, we have sum rules (see, for example, Refs.\cite{Blaizot:2001nr,Aarts:2002tn})
\bear
\int^{|\bm p|}_{-|\bm p|} dp^0 \,\rho^{\rm HTL}_s(p^0,|\bm p|)&=&{\pi\over 2}{ m_f^2\over |\bm p|^2}\left(\log\left({4|\bm p|^2\over m_f^2}\right)-1\right)\,,\nonumber\\
\int^{|\bm p|}_{-|\bm p|} dp^0 \,p^0\,\rho^{\rm HTL}_s(p^0,|\bm p|)&=&s{\pi \over 2}{m_f^2\over |\bm p|^2}\left(\log\left({4|\bm p|^2\over m_f^2}\right)-3\right)\,,
\eear
which gives
\bear
(2\pi)^3 2\omega {d\Gamma^{\rm odd}_{\rm soft}\over d^3\bm k}&\approx&d_R{e^2\over (2\pi)}m_f^2\left(n_+(\omega)n_-(0)-n_-(\omega)n_+(0)\right)\int_{m_f}^{q^*} d|\bm p|\,{1\over |\bm p|}\nonumber\\&=&d_R {e^2\over (2\pi)}m_f^2\left(n_+(\omega)n_-(0)-n_-(\omega)n_+(0)\right)\log(q^*/m_f)\,.
\eear
Looking at the leading log from the hard Compton and Pair Annihilation processes (\ref{LLhard}),
\be
(2\pi)^3 2\omega {d\Gamma^{\rm odd}_{\rm hard}\over d^3\bm k}\approx d_R {e^2\over (2\pi)}m_f^2\left(n_+(|\bm k|)n_-(0)-n_-(|\bm k|)n_+(0)\right)\log\left({T/q^*}\right)\,,
\ee
we see that the $\log(q^*)$ nicely cancels in their sum, which is an important consistency check of our computation.

\subsection{Physics of Leading Log Result\label{subsec3}}

Looking at the leading log expressions for both P-even case (\ref{pevenLO}) and the P-odd emission rate (\ref{poddLO}),
\bear
(2\pi)^3 2\omega {d\Gamma^{\rm total}_{\rm soft}\over d^3\bm k}&\approx& d_R {e^2\over (2\pi)}m_f^2\left(n_+(\omega)n_-(0)+n_-(\omega)n_+(0)\right)\log(q^*/m_f)\,,\nonumber\\
(2\pi)^3 2\omega {d\Gamma^{\rm odd}_{\rm soft}\over d^3\bm k}&\approx& d_R {e^2\over (2\pi)}m_f^2\left(n_+(\omega)n_-(0)-n_-(\omega)n_+(0)\right)\log(q^*/m_f)\,,
\eear
and recalling that they are given in terms of spin polarized emission rates as
\be
\Gamma^{\rm total}=\Gamma(\epsilon^+)+\Gamma(\epsilon^-)\,,\quad \Gamma^{\rm odd}=\Gamma(\epsilon^+)-\Gamma(\epsilon^-)\,,
\ee
we find that the leading log spin polarized emission rates are given, after matching the logarithmic dependence on $q^*$ with the hard rate, as
\be
(2\pi)^3 2\omega {d\Gamma(\epsilon^\pm)\over d^3\bm k}\bigg|_{\rm Leading\,Log}=d_R {e^2\over (2\pi)}m_f^2\,n_\pm(\omega)n_\mp(0)\log(T/m_f)\,,\label{polarizedrate}
\ee
which can be physically understood as follows.

\begin{figure}[t]
 \centering
 \includegraphics[height=5cm]{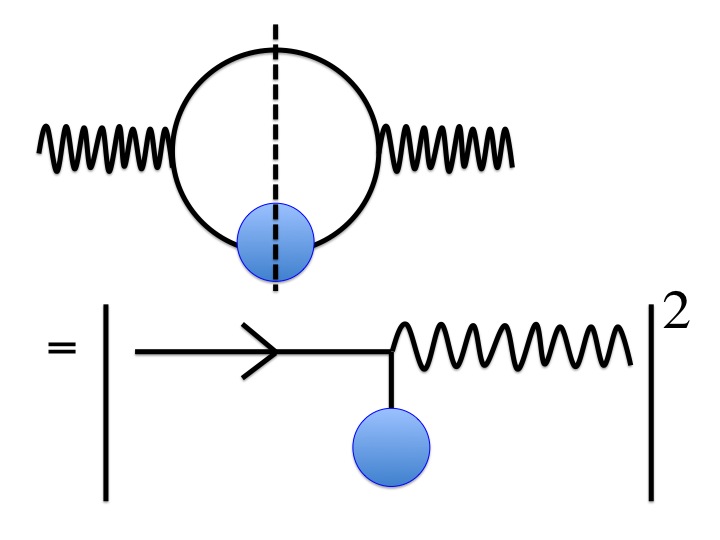}\caption{Leading log contributions from soft t- or u-channel exchanges: a hard fermion making conversion to a collinear photon. The blob represents Hard Thermal Loop (HTL) re-summed propagator. \label{figLL1}}
 \end{figure}
Recall that the leading log comes from the soft t-channel fermion exchange, and the t-channel momentum is space-like as can be seen in
the integral in (\ref{softres}); we have $p^0<|\bm p|$. The spectral density in this kinematics is non-zero due to Landau damping that is captured by HTL self-energy, and represents thermally excited (fermionic) fluctuations of soft momentum  that are present in the finite temperature plasma. The leading log process can be
understood as a process of a hard fermion making conversion into a collinear photon after being annihilated by a soft fermion of momentum $gT$, as in the Figure \ref{figLL1}. At leading order, this $gT$ momentum can be taken as zero.

\begin{figure}[t]
 \centering
 \includegraphics[height=6cm]{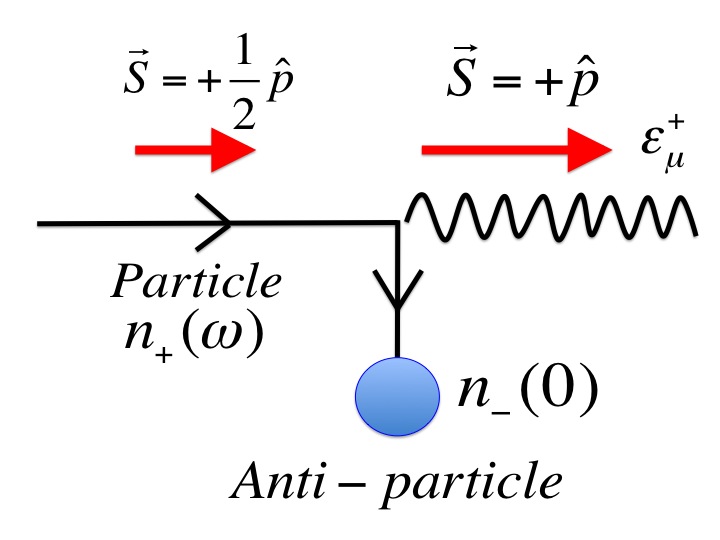}\caption{ Angular momentum conservation in leading log spin polarized emission rates. \label{figLL2}}
 \end{figure}

For definite spin helicity of the final photon in $\Gamma(\epsilon^\pm)$, the conservation of angular momentum dictates that the incoming hard fermion which is collinear to the photon should have a spin $\pm 1/2$ aligned with the momentum direction: the other spin $\pm 1/2$ to make up the final spin $\pm 1$ of the photon will be provided by the annihilating soft fermion.
Since hard fermions have bare spectral density at leading order in coupling, they have definite helicities determined by their quantization in free limit: for our right-handed Weyl fermion field, a particle has helicity $+1/2$ and anti-particle has $-1/2$.
This means that the leading log rate of $\Gamma(\epsilon^+)$ (for photons of spin helicity $+1$) can appear only from the incoming particle of helicity $+1/2$, while an incoming anti-particle of helicity $-1/2$ can not contribute to $\Gamma(\epsilon^+)$. Since the incoming particle can annihilate only with a soft anti-particle, the rate $\Gamma(\epsilon^+)$ should be proportional to $n_+(\omega) n_-(0)$, where the first factor is the number density of incoming particle and the second is the number density of annihilating anti-particle of zero (soft) momentum.
See Figure \ref{figLL2}. The precisely same logic tells us that the leading log rate of $\Gamma(\epsilon^-)$ should be proportional to $n_-(\omega)n_+(0)$.
This argument nicely explains the result in (\ref{polarizedrate}). The overall $m_f^2$ is nothing but the strength of the fermionic spectral density
in soft momentum range that arises from the HTL self-energy: the same self-energy also gives arise to the asymptotic thermal mass $m_f^2$.

\subsection{Collinear Bremstrahlung and Pair Annihilation: LPM Resummation\label{subsec4}}

In this section, we compute collinear Bremstrahlung and Pair Annihilation contributions to the P-odd photon emission rate that are induced by multiple scatterings with soft thermal gluons in the plasma \cite{Arnold:2001ba}. The incoming quark or anti-quark of a hard momentum experiences soft transverse kicks by thermal gluons of momenta $\sim gT$, becoming off-shell by small amount $g^2 T$, during which a nearly collinear photon is emitted, or quark-antiquark pair annihilates to a collinear photon. The rate of these soft scatterings is well-known to be $\sim g^2 T$ (which causes the damping rate of $\sim g^2T$). The scattering gluons are genuine thermal effects: their momenta are space like and the non-zero spectral density in this kinematics arises only due to the Landau damping.  Since the life time of the intermediate states dictated by small virtuality $g^2 T$ is of $1/(g^2T)$, which is comparable to the scattering rate, one has to sum over all multiple scatterings to get the correct leading order result, coined as the LPM re-summation \cite{Arnold:2001ba}. These contributions add to the leading order constant under the log.
The effect of re-summation typically gives a suppression compared to the single scattering contribution.

\begin{figure}[t]
 \centering
 \includegraphics[height=3cm, width=12cm]{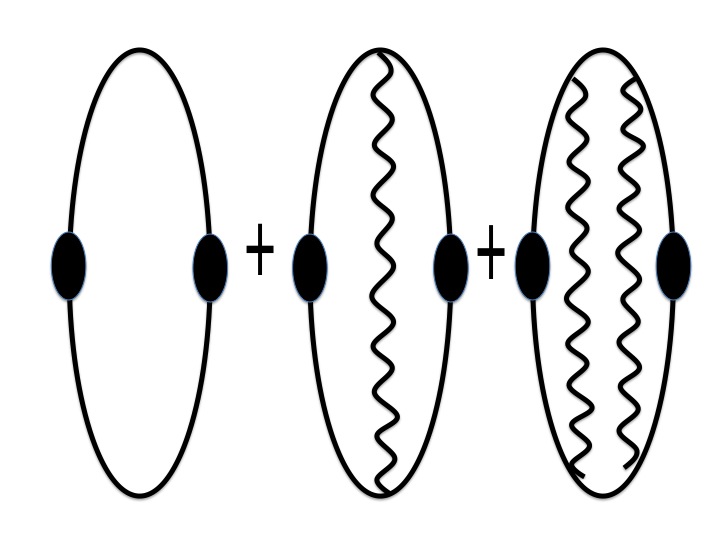}\caption{Ladder diagrams to be summed over to get the correct leading order LPM contribution to (our P-odd) photon emission rate. \label{fig7}}
 \end{figure}

In diagrammatic language, the LPM re-summation corresponds to summing over all ladder diagrams of the type depicted in Figure \ref{fig7} for the retarded (or ``ra'') current-current correlation functions that enter the photon emission rate formula \cite{Arnold:2001ba}.
The reason why these multiple ladder diagrams are
not suppressed by higher powers in coupling constant is the presence of collinear ``pinch'' singularities arising from nearly on-shell fermion propagators: the momentum transfer by exchanged gluon lines are soft, and each pair of fermion propagators, one from the upper line and the other from the lower line, are nearly on-shell and have an IR pinch singularity when the internal momentum is nearly collinear to the external photon momentum (the detail will become clear in the following).
This singularity is regulated by soft transverse component of the fermion momentum, ${\bm p}_\perp^2\sim g^2T^2$, induced by soft kicks from thermal gluons. Then, one has to also include in the propagators the fermion thermal mass $m_f^2\sim g^2T^2$ and the leading order damping rate $\zeta\sim g^2 T$ which are of the same order as ${\bm p}_\perp^2$.

\begin{figure}[t]
 \centering
 \includegraphics[height=7cm]{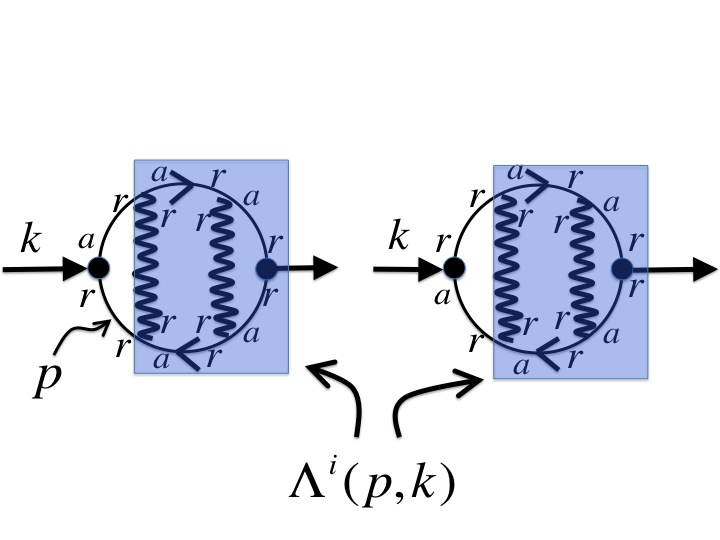}\caption{Two types of real-time ladder diagrams for leading order LPM contributions. The shaded part represents the re-summed rr-type current vertex $\Lambda^i(p,k)$. The rr-type gluon lines are the HTL re-summed ones. \label{fig8}}
 \end{figure}

\begin{figure}[t]
 \centering
 \includegraphics[height=7cm]{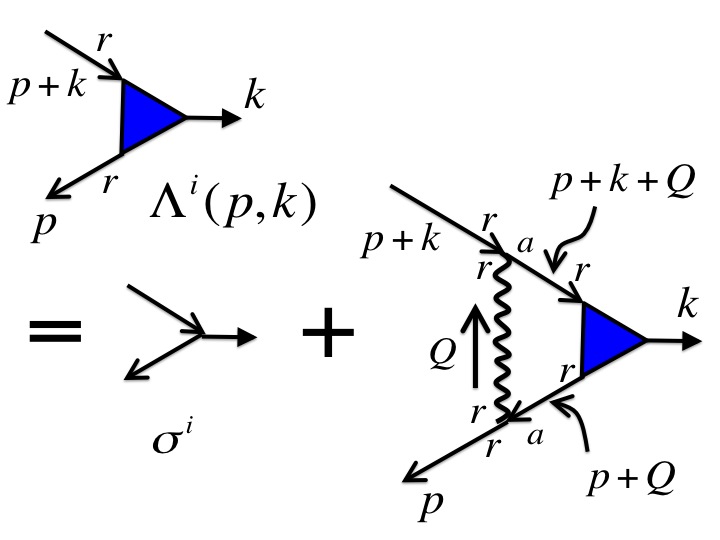}\caption{The real-time Schwinger-Dyson equation for the re-summed vertex $\Lambda^i(p,k)$.  \label{fig9}}
 \end{figure}

Since the exchanged gluons have soft momenta for leading order contributions, we need to re-sum gluonic HTL self-energy
in their propagators.
To get a Bose-Einstein enhancement $n_B(q^0)\sim T/q^0\sim 1/g$ in the exchanged gluon lines, the gluon propagators need to be of the $rr$-type in the ``ra''-basis of Schwinger-Keldysh formalism: only these diagrams give leading order contributions in $g$.
Imposing this requirement and the maximal number of pinch singularities (that arise from a pair of $S^{ra}$ and $S^{ar}$ propagators), there are essentially two types of ladder diagrams to be summed over in the ``ra''-basis as depicted in Figure \ref{fig8}.
Defining the re-summed``rr''-type fermion-current vertex $\Lambda^i(p,k)$ which has two r-type fermions legs, the re-summed
$G^{ra}_{ij}(k)$ current-current correlation function is written as
\be
G^{ra}_{ij}(k)=(-1)d_R\int {d^4 p\over (2\pi)^4}\,{\rm tr}\left[S^{ra}(p+k)\sigma^j S^{rr}(p)\Lambda^i(p,k)+S^{rr}(p+k)\sigma^j S^{ar}(p)\Lambda^i(p,k)\right]\,.
\ee
Since the pinch singularity appears from a pair of $S^{ra}$ and $S^{ar}$, and using the thermal relation $S^{rr}(p)=(1/2-n_+(p^0))(S^{ra}(p)-S^{ar}(p))$, the singular part of $G^{ra}_{ij}(k)$ is given by ($\omega\equiv k^0=|\bm k|$)
\be
G^{ra}_{ij}(k)\approx d_R\int {d^4 p\over (2\pi)^4} \left(n_+(p^0+\omega)-n_+(p^0)\right) {\rm tr}\left[S^{ra}(p+k) \sigma^j S^{ar}(p) \Lambda^i(p,k)\right]\,.\label{pinch}
\ee
The re-summation of the vertex $\Lambda^i(p,k)$ is achieved by solving the Schwinger-Dyson equation described in the Figure \ref{fig9},
\be
\Lambda^i(p,k)=\sigma^i+(ig)^2C_2(R)\int {d^4 Q\over (2\pi)^4}\sigma^\beta S^{ar}(p+Q)\Lambda^i(p+Q,k)S^{ra}(p+Q+k)\sigma^\alpha {\cal G}^{rr}_{\alpha\beta}(Q)\,,\label{lambda}
\ee
where ${\cal G}^{rr}_{\alpha\beta}$ is the HTL re-summed gluon propagator.
We will solve this integral equation and compute $G^{ra}_{ij}(k)$ in leading collinear pinch singularity limit.

The real-time fermion propagators, including the thermal mass and the leading order damping rate, are given as
\be
S^{ra}(p)=\sum_s {i{\cal P}_s(\bm p)\over p^0-s\sqrt{|\bm p|^2+m_f^2}+{i\over 2}\zeta}= -(S^{ar}(p))^\dagger\,,
\ee
where the damping rate is independent of momentum $p$ and the species $s$ at leading order
\be
\zeta =C_2(R){g^2\over 2\pi}\log(1/g)T\,.
\ee
Let's consider the pair of $S^{ra}(p+k)$ and $S^{ar}(p)$ in (\ref{pinch}) to illustrate the pinch singularity and its leading order treatment. Looking at the expression
\be
S^{ra}(p+k)S^{ar}(p)=\sum_{s,t}{i {\cal P}_s(\bm p+\bm k)\over \left(p^0+|\bm k|-s\sqrt{|\bm p+\bm k|^2+m_f^2}+{i\over2}\zeta\right)}\,{i{\cal P}_t(\bm p)\over  \left(p^0-t\sqrt{|\bm p|^2+m_f^2}-{i\over 2}\zeta\right)}\,,\label{pinch2}
\ee
the two poles in the complex $p^0$-plane, one in the upper half plane and the other in the lower half plane,
\be
p^0=-|\bm k|+s\sqrt{|\bm p+\bm k|^2+m_f^2}-{i\over 2}\zeta\,,\quad p^0=t\sqrt{|\bm p|^2+m_f^2}+{i\over 2}\zeta\,,
\ee
may be close to each other with a distance of $\sim g^2 T$, if $\bm p$ is nearly collinear to $\bm k$ and ${\bm p}_\perp\sim gT$.
In computing $p^0$ integral, we close the $p^0$ integral contour, say, in the upper half plane, picking up the pole of $p^0=t\sqrt{|\bm p|^2+m_f^2}+{i\zeta/2}$,
then the residue from the other pole is
\be
{1\over |\bm k|+ t\sqrt{|\bm p|^2+m_f^2}-s\sqrt{|\bm p+\bm k|^2+m_f^2}+i\zeta}\,.
\ee
 Let's fix the direction of $\bm k$ to be along $\hat{\bm z}=\hat{\bm x}^3$ direction, and write the $\hat{\bm z}$ component of momentum $\bm p$ as $p_\pl$, and the perpendicular component as ${\bm p}_\perp$, so that we can expand up to order $g^2 T$ as
\be
\sqrt{|\bm p|^2+m_f^2}\approx  |p_\pl|+{{\bm p}_\perp^2+m_f^2\over 2|p_\pl|}\,,\quad
\sqrt{|\bm p+\bm k|^2+m_f^2}\approx |p_\pl+|\bm k||+{{\bm p}_\perp^2+m_f^2\over 2|p_\pl+|\bm k||}\,.
\ee The pinch singularity happens when the leading collinear terms in the denominator cancel with each other, that is $|\bm k|+t|p_\pl|-s|p_\pl+|\bm k||=0$, to result in $\sim g^2 T$ in the denominator which
enhances the contribution.
There are three physically distinct cases where this happens:

1) $s=t=1$: in this case, $|\bm k|+|p_\pl|-|p_\pl+|\bm k||=0$ is satisfied when $p_\pl>0$. Considering the kinematics, one easily sees that this case corresponds to quark of momentum $\bm p+\bm k$ emitting the collinear photon of momentum $\bm k$ by Bremstrahlung.
The residue becomes
\be
{{\bm p}_\perp^2+m_f^2\over 2 p_\pl}-{{\bm p}_\perp^2+m_f^2\over 2 (p_\pl+|\bm k|)}+i\zeta={|\bm k|({\bm p}_\perp^2+m_f^2)\over 2p_\pl (p_\pl+|\bm k|)}+i\zeta \equiv \delta E({\bm p}_\perp)+i\zeta\,.\label{res}
\ee

2) $s=1, t=-1$:  the condition $|\bm k|-|p_\pl|-|p_\pl+|\bm k||=0$ is fulfilled when $-|\bm k| <p_\pl < 0$, and this case corresponds to collinear pair annihilation of a quark of momentum $\bm p+\bm k$ and an anti-quark of momentum $-\bm p$. Considering signs of $p_\pl$ and $p_\pl+|\bm k|$, one finds that the residue has the precisely the same expression, $\delta E+i\zeta$ with $\delta E$ is defined as above.

3) $s=t=-1$: we have $p_\pl<-|\bm k|$, which corresponds to Bremstrahlung of anti-quark of momentum $\bm p+\bm k$.
Again the residue has the precisely the same form as $\delta E+i\zeta$.

Note that in all three cases, $(s,t)$ are correlated with $p_\pl$ in such a way that $s(p_\pl+|\bm k|)>0$ and $t p_\pl>0$.
Since we only care about the above pinch singularity enhanced contributions, the $(s,t)$ are uniquely chosen for each value of $p_\pl$ as above,
and we consider only these terms in the following.

In leading order treatment, the location of the pole can be approximated as $p^0=t\sqrt{|\bm p|^2+m_f^2}+{i\zeta/2}\approx t| p_\pl|=p_\pl$
in all other places in the integral once the above residues are correctly identified. In summary, we can replace the two poles in (\ref{pinch2}) by
\be
{1\over \left(p^0+|\bm k|-s\sqrt{|\bm p+\bm k|^2+m_f^2}+{i\over2}\zeta\right)}\,{1\over  \left(p^0-t\sqrt{|\bm p|^2+m_f^2}-{i\over 2}\zeta\right)}\to {(2\pi i)\delta(p^0-p_\pl)\over \delta E+i\zeta}\,,
\ee
and depending on the value of $p_\pl\in[-\infty,+\infty]$, the suitable $(s,t)$ as described in the above has to be chosen. For example,
we have for (\ref{pinch}),
\bear
S^{ra}(p+k)\sigma^j S^{ar}(p)&\to & \bigg({\cal P}_+(\bm p+\bm k)\sigma^j {\cal P}_+(\bm p)\Theta(p_\pl)+{\cal P}_+(\bm p+\bm k)\sigma^j {\cal P}_-(\bm p)\Theta(-p_\pl)\Theta(p_\pl+|\bm k|)\nonumber\\
&+&{\cal P}_-(\bm p+\bm k)\sigma^j {\cal P}_-(\bm p)\Theta(-p_\pl-|\bm k|)\bigg)\,\,{-(2\pi i)\delta(p^0-p_\pl)\over \delta E({\bm p}_\perp)+i\zeta}\,.\label{pinch3}
\eear

Since $Q$ carried by exchange gluons is soft, we have an essentially same structure for $S^{ar}(p+Q)\Lambda^i(p+Q,k)S^{ra}(p+Q+k)$ appearing in the integral equation for $\Lambda^i(p,k)$ in (\ref{lambda}),
\bear
&&S^{ra}(p+Q)\Lambda^i(p+Q,k) S^{ar}(p+Q+k)\nonumber\\&\to & \bigg({\cal P}_+(\bm p+\bm q)\Lambda^i(p+Q,k) {\cal P}_+(\bm p+\bm q+\bm k)\Theta(p_\pl)\nonumber\\&+&{\cal P}_-(\bm p+\bm q)\Lambda^i(p+Q,k) {\cal P}_+(\bm p+\bm q+\bm k)\Theta(-p_\pl)\Theta(p_\pl+|\bm k|)\nonumber\\
&+&{\cal P}_-(\bm p+\bm q)\Lambda^i(p+Q,k){\cal P}_-(\bm p+\bm q+\bm k)\Theta(-p_\pl-|\bm k|)\bigg)\,\,{-(2\pi i)\delta(q^0-q_\pl)\over \delta E({\bm p}_\perp+{\bm q}_\perp)+i\zeta}\,,\nonumber\\\label{pinch4}
\eear
the only difference of which are the argument ${\bm p}_\perp+{\bm q}_\perp$ in $\delta E$ instead of ${\bm p}_\perp$.
In writing the $\delta(q^0-q_\pl)$ factor, we used $p^0=p_\pl$ that is imposed by (\ref{pinch3}) when we compute the correlation function $G^{ra}_{ij}(k)$ by (\ref{pinch}). We will solve the integral equation (\ref{lambda}) for $\Lambda^i$, with the above replacement (\ref{pinch4}) that is enough for the leading order result.

Looking at (\ref{pinch}), (\ref{pinch3}), and (\ref{pinch4}), what we need are the projected vertices
\be
{\cal P}_s(\bm p+\bm k)\sigma^j{\cal P}_t(\bm p)\equiv \Sigma^j_{st}(\bm p,\bm k)\,{\cal P}_s(\bm p+\bm k){\cal P}_t(\bm p)\,,\label{Sig}
\ee
and we define a vector function $F^i({\bm p}_\perp)$ as (we ignore $p_\pl$ and $|\bm k|$ arguments in $F^i$ as they are common in all subsequent expressions)
\be {\cal P}_t(\bm p)\Lambda^i(p,k)\big|_{p^0=p_\pl}{\cal P}_s(\bm p+\bm k)\equiv \left(\delta E({\bm p}_\perp)+i\zeta\right)\,F^i({\bm p}_\perp)\,\,{\cal P}_t(\bm p){\cal P}_s(\bm p+\bm k)\,.\label{Fi}
\ee
Here, we emphasize again that the $(s,t)$ are the choice depending on the value of $p_\pl$ suitable for the pinch singularity that we discuss in the above.
Note that $\Sigma^j_{st}$ and $F^i$ are complex valued functions, not $2\times2$ matrices.
In terms of these functions, using (\ref{pinch}), (\ref{pinch3}), (\ref{Sig}) and (\ref{Fi}), we have (recall $\omega\equiv k^0=|\bm k|$)
\bear
G^{ra}_{ij}(k)&=&d_R(-i)\int {d^4 p\over (2\pi)^4}(n_+(p^0+\omega)-n_+(p^0)) \Sigma^j_{st}(\bm p,\bm k)F^i({\bm p}_\perp){\rm tr}\left({\cal P}_s(\bm p+\bm k){\cal P}_t(\bm p)\right)\nonumber \\ &\times&(2\pi)\delta(p^0-p_\pl)\nonumber \\ &\approx&
d_R(-i)\int {d^4 p\over (2\pi)^4}(n_+(p^0+\omega)-n_+(p^0)) \Sigma^j_{st}(\bm p,\bm k)F^i({\bm p}_\perp)(2\pi)\delta(p^0-p_\pl)\,,\nonumber\\\label{gra}
\eear
where in the last line, we use
\be
{\rm tr}\left({\cal P}_s(\bm p+\bm k){\cal P}_t(\bm p)\right)={1\over 2}\left({1+ st \,\hat{\bm p}\cdot\widehat{\bm p+\bm k}}\right)\approx 1\,,
\ee
to leading order in ${\bm p}_\perp/p_\pl\sim g$ and we use $tp_\pl>0$ and $s(p_\pl+|\bm k|)>0$.

Recall that our P-odd photon emission rate is given in terms of $G^{ra}_{ij}(k)$ as
\be
(2\pi)^3 2\omega {d\Gamma^{\rm odd}\over d^3 \bm k}=e^2 n_B(\omega)(-2){\rm Im}\left[G^{ra}_{12}(k)-G^{ra}_{21}(k)\right]\,,
\ee
given the choice of $\bm k=|\bm k|\hat{\bm x}^3$. Hence, we need only the transverse components of $\Sigma^j_{st}$ and $F^i$.
A short computation from the definition (\ref{Sig}) after taking the trace of the both sides gives
\be
\Sigma^j_{st}(\bm p,\bm k)={s\,\widehat{\bm p+\bm k}^j+t\,\hat{\bm p}^j+i st \,\epsilon^{jlm}\hat{\bm p}^l\,\widehat{\bm p+\bm k}^m\over
1+st\,\hat{\bm p}\cdot\widehat{\bm p+\bm k}}\,,\label{sigexp}
\ee
and the integral equation (\ref{lambda}) after being contracted with ${\cal P}_t(\bm p)$ on the left and ${\cal P}_s(\bm p+\bm k)$
on the right gives
\be
\left(\delta E({\bm p}_\perp)+i\zeta\right)F^i({\bm p}_\perp)=\left(\Sigma^i_{st}(\bm p,\bm k)\right)^*+g^2C_2(R)\int {d^4 Q\over (2\pi)^4}F^i({\bm p}_\perp+{\bm q}_\perp) \hat v^\alpha \hat v^\beta {\cal G}^{rr}_{\alpha\beta}(Q)
(2\pi i)\delta(q^0-q_\pl)\,,\label{integ}\ee
where in the integral kernel, we used an approximation
\be
{\cal P}_t(\bm p)\sigma^\beta{\cal P}_t(\bm p+\bm q)\approx {\cal P}_t(\bm p)\sigma^\beta{\cal P}_t(\bm p)=p^\beta_t/|p_\pl|{\cal P}_t(\bm p)\,,\label{Q}
\ee
for soft $Q$, where $p^\alpha_t=(|\bm p|,t\bm p)\approx (|p_\pl|, 0,0,t p_\pl)$ at leading order, so that
$p^\alpha_t/|p_\pl|$ is a light-like 4-velocity $\hat v^\alpha$ along the collinear vector $t\bm p$. Considering the correlation between $p_\pl$ and the sign of $t$ that we describe before, we see that $t p_\pl>0$ always, so that this 4-velocity is always $\hat v^\alpha=(1,0,0,1)$. The same is true for ${\cal P}_s(\bm p+\bm q+\bm k)\sigma^\alpha{\cal P}_s(\bm p+\bm k)$ so that we have
\be
{\cal P}_t(\bm p)\sigma^\beta{\cal P}_t(\bm p+\bm q){\cal P}_s(\bm p+\bm q+\bm k)\sigma^\alpha{\cal P}_s(\bm p+\bm k)\approx
\hat v^\alpha \hat v^\beta{\cal P}_t(\bm p){\cal P}_s(\bm p+\bm k)\,,
\ee
which has been used to arrive at our integral equation for $F^i$ in (\ref{integ}). Since $F^i\sim 1/g$ and the both sides of (\ref{integ}) are of order $\sim g$, this approximation is enough for the leading order computation.

One subtle point is that the HTL gluon fluctuations in ${\cal G}^{rr}_{\alpha\beta}$ contains a P-odd spectral density\footnote{See the appendices in Ref.\cite{Jimenez-Alba:2015bia} for some of its sum rules in the HTL approximation.}
which is anti-symmetric in $\alpha$ and $\beta$, which could potentially contribute to our P-odd photon emission rate, if we keep $Q$ corrections in (\ref{Q}). We estimated them to find that these corrections are higher order in $g$.
The fluctuations contracted with light-like vector $\hat v^\alpha$ in (\ref{integ}), $\hat v^\alpha \hat v^\beta {\cal G}^{rr}_{\alpha\beta}$ (which are the correlations along the Eikonalized light-like
Wilson line) receive only the usual P-even longitudinal and transverse contributions.

As is well-known \cite{Arnold:2001ba}, the integral equation is further simplified due to the fact that the integral on the right  in (\ref{integ}) without $F^i$ is identical to the leading order damping rate~$\zeta$,
\be
\zeta=g^2C_2(R)\int {d^4 Q\over (2\pi)^4}\hat v^\alpha \hat v^\beta {\cal G}^{rr}_{\alpha\beta}(Q)
(2\pi)\delta(q^0-q_\pl)\,,
\ee
so that we can move $i\zeta\, F^i({\bm p}_\perp)$ term in the left to the right to arrive at
\bear
\delta E({\bm p}_\perp)F^i({\bm p}_\perp)&=&\left(\Sigma^i_{st}(\bm p,\bm k)\right)^*\\&+&g^2C_2(R)\int {d^4 Q\over (2\pi)^4}\left(F^i({\bm p}_\perp+{\bm q}_\perp)-F^i({\bm p}_\perp)\right) \hat v^\alpha \hat v^\beta {\cal G}^{rr}_{\alpha\beta}(Q)
(2\pi i)\delta(q^0-q_\pl)\,.\nonumber
\eear
This form has a good infrared behavior so that only the well-controlled soft scale $Q\sim gT$ contributes at leading order, while the magnetic scale of $g^2 T$ gives a finite, sub-leading contributions.

Finally, for soft $Q$ we replace
\be
{\cal G}^{rr}_{\alpha\beta}(Q)= \left({1\over 2}+n_B(q^0)\right)\rho^{\rm gluon}_{\alpha\beta}(Q)\approx {T\over q^0}\,\rho^{\rm gluon}_{\alpha\beta}(Q)\,,
\ee
for leading order, where $\rho^{\rm gluon}_{\alpha\beta}$ is the gluon spectral density in HTL approximation, and the amazing sum rule in Ref.\cite{Aurenche:2002pd} gives the integral over $(q^0,q_\pl)$ as
\be
T\int {dq^0 dq_\pl\over (2\pi)^2} \hat v^\alpha \hat v^\beta{1\over q^0} \rho^{\rm gluon}_{\alpha\beta}(Q) (2\pi)\delta(q^0-q_\pl)={T m_D^2\over {\bm q}_\perp^2 ({\bm q}_\perp^2+m_D^2)}\,,
\ee
where
\be
m_D^2={g^2}\left({T^2\over 3}+{\mu^2\over\pi^2}\right)(T_A+N_F T_R)={g^2}\left({ T^2\over 3}+{\mu^2\over\pi^2}\right)(N_c+N_F/2)\,,
\ee is the Debye mass for $N_F$ Dirac quarks in fundamental representation, so that the integral equation for $F^i({\bm p}_\perp)$ is finally recast to
\be
\delta E({\bm p}_\perp)F^i({\bm p}_\perp)=\left(\Sigma^i_{st}(\bm p,\bm k)\right)^*+i\int {d^2 {\bm q}_\perp\over (2\pi)^2}\,{\cal C}({\bm q}_\perp)\,\left(F^i({\bm p}_\perp+{\bm q}_\perp)-F^i({\bm p}_\perp)\right)\,,\label{intfin}
\ee
with \be C({\bm q}_\perp)=g^2 C_2(R){T\, m_D^2\over {\bm q}_\perp^2 ({\bm q}_\perp^2+m_D^2)}\,.\ee

Since we need only the transverse parts of ({\ref{intfin}) and (\ref{gra}) for $G^{ra}_{ij}(k)$, we expand $\Sigma^{i}_{st}(\bm p,\bm k)$ given in (\ref{sigexp})
to linear order in ${\bm p}_\perp/p_\pl\sim g$, which is enough for leading order,
\bear
\Sigma^i_{st}(\bm p,\bm k)&\approx& {1\over 2}\left({1\over p_\pl}+{1\over p_\pl+|\bm k|}\right){\bm p}_\perp^i+{i\over 2}\left({1\over p_\pl}-{1\over p_\pl+|\bm k|}\right)\epsilon_\perp^{il}\,{\bm p}_\perp^l\nonumber\\
&=&{2p_\pl+|\bm k|\over 2p_\pl(p_\pl+|\bm k|)}{\bm p}_\perp^i+i{|\bm k|\over 2p_\pl(p_\pl+|\bm k|)}\epsilon_\perp^{il}\,{\bm p}_\perp^l\,,
\eear
where we used the fact that $t p_\pl>0$ and $s(p_\pl+|\bm k|)>0$, and $\epsilon_\perp^{12}=-\epsilon_\perp^{21}=1$. We use this expansion in both (\ref{gra}) and (\ref{intfin}). From (\ref{intfin}), we see that the solution for $F^i({\bm p}_\perp)$ is given by
\be
F^i({\bm p}_\perp)={2p_\pl+|\bm k|\over 2p_\pl(p_\pl+|\bm k|)} f^i_\perp({\bm p}_\perp)-i{|\bm k|\over 2p_\pl(p_\pl+|\bm k|)}\epsilon_\perp^{il}\,f^l_\perp({\bm p}_\perp)\,,
\ee
where $f^i_\perp({\bm p}_\perp)$ is the solution of the integral equation
\be
\delta E({\bm p}_\perp)f^i_\perp({\bm p}_\perp)={\bm p}_\perp^i+i\int {d^2 {\bm q}_\perp\over (2\pi)^2}\,{\cal C}({\bm q}_\perp)\,\left(f^i_\perp({\bm p}_\perp+{\bm q}_\perp)-f^i_\perp({\bm p}_\perp)\right)\,.\label{amy}
\ee
This equation for $f^i_\perp({\bm p}_\perp)$ is identical to the integral equation obtained by Arnold-Moore-Yaffe in Ref.\cite{Arnold:2001ba}, with the identification
\be
f^i_\perp({\bm p}_\perp)=-{i\over 2} \left(f^i_{\rm AMY}({\bm p}_\perp)\right)^*\,,
\ee
so that the techniques of solving this integral equation that are known in literature can be utilized to find our object $F^i({\bm p}_\perp)$.
Using this expression for $F^i$ and (\ref{gra}) for $G^{ra}_{ij}(k)$, we obtain after short manipulations,
\be
G^{ra}_{12}-G^{ra}_{21}(k)=-{d_R\over 2}\int {dp_\pl d^2 {\bm p}_\perp \over (2\pi)^3}
(n_+(p_\pl+\omega)-n_+(p_\pl)){|\bm k|(2p_\pl+|\bm k|)\over p_\pl^2(p_\pl+|\bm k|)^2} ({\bm p}_\perp\cdot {\bm f}_\perp)\,,
\ee
and using an interesting identity
\be
n_B(\omega)\left(n_+(p_\pl+\omega)-n_+(p_\pl)\right)=-n_+(p_\pl+\omega) \left(1-n_+(p_\pl)\right)\,,
\ee
we finally arrive at an expression for our P-odd photon emission rate in terms of the solution ${\bm f}_\perp({\bm p}_\perp)$ of the integral equation (\ref{amy}) (recall $\omega=|\bm k|$),
\be
(2\pi)^3 2\omega {d\Gamma^{\rm odd}_{\rm LPM}\over d^3\bm k}=e^2 d_R
\int {dp_\pl d^2 {\bm p}_\perp \over (2\pi)^3}
n_+(p_\pl+\omega) \left(1-n_+(p_\pl)\right){\omega(2p_\pl+\omega)\over p_\pl^2(p_\pl+\omega)^2} (-1){\rm Im}\left[({\bm p}_\perp\cdot {\bm f}_\perp) \right]\,.\label{lpmfinal}
\ee
This is the main outcome of this section. Our numerical evaluation is based on this expression with the integral equation (\ref{amy}), where $\delta E$ is given in (\ref{res}) (see also (\ref{deltaE})).

Although it is not manifestly obvious that the above expression is an odd function in (axial) chemical potential $\mu$ that enters the distribution function $n_+$, one way to see this is to first observe that the factor $n_+(p_\pl+\omega) \left(1-n_+(p_\pl)\right)$ is easily recognized as the statistical factor for the collinear Bremstrahlung process of a fermion of momentum ${\bm p}+\bm k$ emitting a photon of momentum $\bm k$, {\it provided} that
$p_\pl>0$. In the case $p_\pl<-|\bm k|$, using the identity
\be
n_+(p_\pl+\omega) \left(1-n_+(p_\pl)\right)=n_-(-p_\pl) \left(1-n_-(-p_\pl-\omega)\right)
\ee
we see that the process is in fact the Bremstrahlung of anti-fermion of momentum $-\bm p$ emitting a photon of momentum $\bm k$.
It is more convenient to change the integration variable in this case to $p_\pl\to -(\tilde p_\pl+\omega)$ so that we have $\tilde p_\pl>0$ and
the statistical factor becomes
\be
n_-(\tilde p_\pl+\omega) \left(1-n_-(\tilde p_\pl)\right)\,,
\ee
which makes the interpretation clearer.
From the expression for $\delta E$ in (\ref{res}), we have
\be
\delta E={\omega({\bm p}_\perp^2+m_f^2)\over 2p_\pl (p_\pl+\omega)}={\omega({\bm p}_\perp^2+m_f^2)\over 2\tilde p_\pl (\tilde p_\pl+\omega)}\,,\label{deltaE}
\ee
so that the integral equation (\ref{amy}) and hence the solution ${\bm f}_\perp({\bm p}_\perp)$ is invariant under this change of variable, but
the integral kernel in our P-odd emission rate in (\ref{lpmfinal}) changes sign under this transformation as
\be
{\omega(2p_\pl+\omega)\over p_\pl^2(p_\pl+\omega)^2}\to -{\omega(2\tilde p_\pl+\omega)\over \tilde p_\pl^2(\tilde p_\pl+\omega)^2}
\,,
\ee
so that the net sign of the contribution from anti-fermion Bremstrahlung is opposite to the one from fermion Bremstrahlung. This is expected since fermion and anti-fermion from our right-handed Weyl fermion field have opposite chirality, so their contributions to $\Gamma^{\rm odd}$ should be opposite. From the above, if we sum over $p_\pl>0$ and $\tilde p_\pl>0$ regions (and calling $\tilde p_\pl$ as $p_\pl$), we see that the final result is proportional to
\be
n_+(p_\pl+\omega) \left(1-n_+(p_\pl)\right)-n_-(p_\pl+\omega) \left(1-n_-( p_\pl)\right)\,,
\ee
which is indeed an odd function on the (axial) chemical potential $\mu$. More generally, by the change of variable from $p_\pl$ to $\tilde p_\pl$ for the entire range of $p_\pl$, we can simply replace the statistical factor in our main formula (\ref{lpmfinal}) with
the average
\be
n_+(p_\pl+\omega) \left(1-n_+(p_\pl)\right)\to {1\over 2}\left(n_+(p_\pl+\omega) \left(1-n_+(p_\pl)\right)-n_-(p_\pl+\omega) \left(1-n_-( p_\pl)\right)\right)\,,
\ee
so that the LPM contribution to our P-odd emission rate, (\ref{lpmfinal}), is now manifestly an odd function in $\mu$.

Following Ref.\cite{Aurenche:2002wq}, the integral equation (\ref{amy}) can be transformed to the one in the transverse 2-dimensional coordinate space ${\bm b}$, which takes a form
\be
{\omega(-\nabla_{\bm b}^2+m_f^2)\over 2 p_\pl(p_\pl+\omega)}{\bm f}^i_\perp({\bm b})=-i{\bm\nabla}_{\bm b}^i \delta^{(2)}({\bm b})+i\, {\cal C}({\bm b}) \,{\bm f}_\perp^i({\bm b})\,,
\ee
where
\be
{\bm f}_\perp^i({\bm b})= \int {d^2 {\bm p}_\perp\over (2\pi)^2}\,e^{i{\bm b}\cdot {\bm p}_\perp}\, f_\perp^i({\bm p}_\perp)\,,
\ee
and
\be
{\cal C}({\bm b})\equiv \int {d^2{\bm q}_\perp\over (2\pi)^2}\,{\cal C}({\bm q}_\perp)\,\left(e^{-i{\bm b}\cdot {\bm q}_\perp}-1\right)=
-{g^2 C_2(R) T\over 2\pi}\left(K_0(|{\bm b}|m_D)+\gamma_E+\log(|{\bm b}| m_D/2)\right)\,.
\ee
From rotational symmetry, one can write
\be
{\bm f}_\perp({\bm b})={\bm b} f(b)\,,\quad b\equiv |\bm b|\,,
\ee
in terms of a scalar function $f(b)$ which satisfies the following second order differential equation
\be
{\omega\over 2p_\pl (p_\pl+\omega)}\left(-\partial_b^2-{3\over b}\partial_b +m_f^2\right)\,f(b)=i \,{\cal C}(b)\, f(b)\,,\label{diff2}
\ee
with the boundary conditions
\be
f(b\to 0)=-i\,{p_\pl (p_\pl+\omega)\over \pi \omega b^2}+{\cal O}(b^0)\,,\quad f(b\to\infty) =0\,.\label{bcd}
\ee
In terms of the scalar function $f(b)$ which can be easily solved from the above differential equation, the ${\bm p}_\perp$ integral in our P-odd emission rate (\ref{lpmfinal})
takes a simple form
\be
\int{d^2{\bm p}_\perp\over (2\pi)^2}\,(-1){\rm Im}\left[{\bm p}_\perp\cdot {\bm f}_\perp({\bm p}_\perp) \right]=(-1){\rm Im}\left[(-i){\bm\nabla}_{\bm b}\cdot{\bm f}_\perp({\bm b})\right]\bigg|_{\bm b\to 0}=2\,{\rm Re}\,f(0)\,,
\ee
so that the final expression for the LPM contribution to the P-odd photon emission rate becomes
\bear
(2\pi)^3 2\omega {d\Gamma^{\rm odd}_{\rm LPM}\over d^3\bm k}&=&e^2 d_R
\int_{-\infty}^{+\infty} {dp_\pl  \over 2\pi}
\left(n_+(p_\pl+\omega) \left(1-n_+(p_\pl)\right)-n_-(p_\pl+\omega) \left(1-n_-(p_\pl)\right)\right)\nonumber\\&\times&{\omega(2p_\pl+\omega)\over p_\pl^2(p_\pl+\omega)^2}\,{\rm Re}\,f(0)\,.\label{lpmfinal2}
\eear
This is what we practically use for numerical evaluations, and the computation reduces to solving the second order differential equation (\ref{diff2}) with the boundary conditions (\ref{bcd}).

\section{Summary of Final Result and Discussion\label{sec4}}

In summary, the leading order P-odd photon emission rate for a single species of right-handed Weyl fermion is a sum of the three contributions: 1) hard Compton and Pair Annihilation rate given by (in t-channel parametrization) the equation (\ref{finalhard10}) with (\ref{A}) where one has to use (\ref{tcq}), 2) soft t- and u-channel contributions given in (\ref{poddLO}), 3) the LPM re-summed collinear Bremstrahlung and Pair Annihilation contribution given in (\ref{lpmfinal2}) with (\ref{diff2}) and (\ref{bcd}).
For a theory with $N_F$ Dirac fermions with an axial chemical potential $\mu_A$, one has to multiply the above results by a factor
\be
2\left(\sum_F Q_F^2\right)\,,
\ee
with a replacement $\mu\to \mu_A$ in the distribution functions, where $Q_F$ are electromagnetic charges of flavor $F$ in units of $e$. Recall also that the Debye mass
\be
m_D^2={g^2}\left({ T^2\over 3}+{\mu^2\over\pi^2}\right)(N_c+N_F/2)\,,
\ee has to be adjusted according to the number of flavors $N_F$.

We choose to present our result in a way similar to the existing literature. Define
\be
{\cal A}(\omega)\equiv 2\,\alpha_{\rm EM} \left(\sum_F Q_F^2\right)d_R {m_{f,(0)}^2\over\omega}n_f(\omega)\,,
\ee
where $n_f(\omega)$ is the Fermi-Dirac distribution with zero chemical potential and $m_{f,(0)}^2\equiv C_2(R)g^2T^2/4$ is the asymptotic fermion thermal mass at zero chemical potential that has to be compared to the full expression (\ref{fmass}) in the presence of (axial) chemical potential
\be
m_f^2=C_2(R) {g^2\over 4}\left(T^2+{\mu_A^2\over\pi^2}\right)\,.
\ee
The hard Compton and Pair Annihilation rate is then written as
\be
(2\pi)^3 {d\Gamma^{\rm odd}_{\rm hard}\over d^3\bm k}={\cal A}(\omega)
{2\over (2\pi)^3}{T\over \omega}{1\over n_f(\omega)}\int^\infty_{q^*} {d|\bm q|\over T}\int^{|\bm q|}_{{\rm max}(-|\bm q|,|\bm q|-2|\bm k|)}{d q^0\over T}\int^{\infty}_{{|\bm q|-q^0\over 2}}{d|\bm p'|\over T}\int^{2\pi}_0 d\phi\,\,\bar{\cal I}\,,\label{finalhard11}
\ee
where
\bear
\bar{\cal I}&=& \left(-{u\over t}-2(t-u)\left({{\bm q}_\perp^2\over t^2}-{{\bm q}_\perp\cdot{\bm p}_\perp'\over tu}\right)\right)\nonumber\\&\times&(n_+(q^0+|\bm k|) n_-(|\bm p'|)-n_-(q^0+|\bm k|)n_+(|\bm p'|))(1+n_B(q^0+|\bm p'|))\nonumber\\ &+& (s-t)\left({1\over t}+{1\over s}-2\left({{\bm q}_\perp\over t}+{({\bm q}_\perp+{\bm p}_\perp')\over s}\right)^2\right)\nonumber\\
&\times&\left(
n_+(q^0+|\bm k|)(1-n_+(q^0+|\bm p'|)) -n_-(q^0+|\bm k|)(1-n_-(q^0+|\bm p'|))\right)n_B(|\bm p'|)\,.\nonumber\\\label{Abar}
\eear
Note that what is multiplied to ${\cal A}(\omega)$ is a dimensionless function on $\omega/T$ (recall $|\bm k|=\omega$), and the phase space integral as well as the integrand $\bar{\cal I}$
is in terms of dimensionless variables $|\bm q|/T$, etc. The soft t- and u-channel contribution is written as
\be
(2\pi)^3 {d\Gamma^{\rm odd}_{\rm soft}\over d^3\bm k}={\cal A}(\omega)
{m_f^2\over m_{f,(0)}^2}{1\over n_f(\omega)}\left(n_+(\omega)n_-(0)-n_-(\omega)n_+(0)\right)\left(\log(q^*/m_f)-1+\log 2\right)\,.
\ee
Finally, the LPM contribution is
\bear
(2\pi)^3 {d\Gamma^{\rm odd}_{\rm LPM}\over d^3\bm k}&=&{\cal A}(\omega){1\over n_f(\omega)}
\int_{-\infty}^{+\infty}d\bar p_\pl\left(n_+(p_\pl+\omega) \left(1-n_+(p_\pl)\right)-n_-(p_\pl+\omega) \left(1-n_-(p_\pl)\right)\right)\nonumber\\&\times&
{\bar\omega(2\bar p_\pl+\bar\omega)\over \bar p_\pl^2(\bar p_\pl+\bar\omega)^2}\,\,{\rm Re}\,\bar f (0)\,,
\eear
where $\bar p_\pl\equiv p_\pl/T$ and $\bar\omega\equiv \omega/T$, and $\bar f(\bar b)$ is the solution of the differential equation
\be
{\bar\omega\over 2\bar p_\pl (\bar p_\pl+\bar\omega)}\left(-\partial^2_{\bar b}-{3\over\bar b}\partial_{\bar b}
+{m_f^2\over m_D^2}\right)\bar f(\bar b)=-i{2\over \pi}{m^2_{f,(0)}\over m_D^2}\left(K_0(\bar b)+\gamma_E+\log(\bar b/2)\right) \bar f(\bar b)\,,
\ee
 with the boundary conditions
 \be
 \bar f(\bar b\to 0)=-i{\bar p_\pl (\bar p_\pl+\bar\omega)\over\pi\bar\omega\bar b^2}{m_D^2\over m_{f,(0)}^2}\,,\quad \bar f(\bar b\to\infty)=0\,.
 \ee

 The final result can be recast to the form
 \be
 (2\pi)^3 {d\Gamma^{\rm odd}_{\rm LO}\over d^3\bm k}={\cal A}(\omega)\left(C^{\rm odd}_{\rm Log}(\omega/T)\log\left(T/ m_f\right)+C^{\rm odd}_{2\leftrightarrow 2}(\omega/T)+C^{\rm odd}_{\rm LPM}(\omega/T)\right)\,,
 \ee
with the dimensionless functions $C^{\rm odd}_{\rm Log}$, $C^{\rm odd}_{2\leftrightarrow 2}$, $C^{\rm odd}_{\rm LPM}$, where
\bear
C^{\rm odd}_{\rm Log}&=&{m_f^2\over m_{f,(0)}^2}{1\over n_f(\omega)}\left(n_+(\omega)n_-(0)-n_-(\omega)n_+(0)\right)\,,\nonumber\\
C^{\rm odd}_{2\leftrightarrow 2}&=&\lim_{q^*\to 0}\bigg({2\over (2\pi)^3}{T\over \omega}{1\over n_f(\omega)}\int^\infty_{q^*} {d|\bm q|\over T}\int^{|\bm q|}_{{\rm max}(-|\bm q|,|\bm q|-2|\bm k|)}{d q^0\over T}\int^{\infty}_{{|\bm q|-q^0\over 2}}{d|\bm p'|\over T}\int^{2\pi}_0 d\phi\,\,\bar{\cal I}\nonumber\\
&-& C^{\rm odd}_{\rm Log}(\omega/T)\left(\log(T/q^*)+1-\log 2\right)\bigg)\,,\nonumber\\
C^{\rm odd}_{\rm LPM}&=&{1\over n_f(\omega)}
\int_{-\infty}^{+\infty}d\bar p_\pl\left(n_+(p_\pl+\omega) \left(1-n_+(p_\pl)\right)-n_-(p_\pl+\omega) \left(1-n_-(p_\pl)\right)\right)\nonumber\\&\times&
{\bar\omega(2\bar p_\pl+\bar\omega)\over \bar p_\pl^2(\bar p_\pl+\bar\omega)^2}\,\,{\rm Re}\,\bar f (0)\,.
\eear
Note that we have not extracted out the energy logarithm given in (\ref{energy}), but one could choose to do so to redefine $C^{\rm odd}_{2\leftrightarrow 2}$.

The above result is valid for full dependence in the axial chemical potential $\mu_A$, but we will present our numerical evaluations only for its linear dependency by expanding the dimensionless functions $C^{\rm odd}_{\rm Log}$, $C^{\rm odd}_{2\leftrightarrow 2}$, $C^{\rm odd}_{\rm LPM}$ in linear order in $\mu_A/T$. In this case, $m_f^2$ can be identified with $m_{f,(0)}^2$ and one can also neglect $\mu_A^2$ in the Debye mass $m_D^2$.
Writing this linear expansion as
\bear
 (2\pi)^3 {d\Gamma^{\rm odd}_{\rm LO}\over d^3\bm k}&\approx&{\cal A}(\omega)\left( C^{\rm odd,(1)}_{\rm Log}(\omega/T)\log\left(T/ m_f\right)+C^{\rm odd,(1)}_{2\leftrightarrow 2}(\omega/T)+C^{\rm odd,(1)}_{\rm LPM}(\omega/T)\right){\mu_A\over T}\nonumber\\
 &+& \mathcal{O}(\mu_A^3)\,,
 \eear
we have
\be
 C^{\rm odd,(1)}_{\rm Log}={1\over 2}\left(1-2n_f(\omega)\right)\,,
 \ee
 while the other two functions, $C^{\rm odd,(1)}_{2\leftrightarrow 2}$, $C^{\rm odd,(1)}_{\rm LPM}$, have to be evaluated numerically.
 The numerical evaluation involves three dimensional integrals and solving second order differential equation, and can be performed with a reasonable precision using Mathematica.
 We present our numerical results in Figure \ref{figFin} for the range $0.5<\omega/T<3$. We see that the LPM contributions to the constant under the log
 is 2-3 times bigger than the one from $2\leftrightarrow 2$ Compton and Pair Annihilation contributions in this range, but we should remember that the leading log contribution comes from these $2\leftrightarrow 2$ processes.

 Finally, recalling that
 \be
\Gamma^{\rm total}=\Gamma(\epsilon^+)+\Gamma(\epsilon^-)\,,\quad \Gamma^{\rm odd}=\Gamma(\epsilon^+)-\Gamma(\epsilon^-)\,,
 \ee
 we get
 \be
 (2\pi)^3 {d\Gamma^{\rm total}_{\rm LO}\over d^3\bm k}\approx{\cal A}(\omega)\left(\log\left(T/ m_f\right)+C^{\rm total,(0)}_{2\leftrightarrow 2}(\omega/T)+C^{\rm total,(0)}_{\rm LPM}(\omega/T)\right)+\mathcal{O}(\mu_A^2)\,,
 \ee
 where
 \bear
	%\label{c2to2fit}
	C^{\rm total,(0)}_{2\leftrightarrow 2}(\omega/T)&=&\frac12
	\ln\left(\frac{2\omega}{T}\right)+ 0.041\frac{T}{\omega} -0.3615 + 1.01 e^{-1.35 \omega/T}\,,\nonumber\\
&& 0.2<\frac{\omega}{T}\,,\\
C^{\rm total,(0)}_{\rm LPM}(\omega/T)
	&=&
	2\left[\frac{0.316\ln(12.18+T/\omega)}{\left(\omega/T\right)^{3/2}}
	+\frac{0.0768\omega/T}{\sqrt{1+\omega/(16.27 T)}}\right],\nonumber\\
    && 0.2<\frac{\omega}{T}<50\,,
	%&&\label{ccolllofit}
\eear
which is nothing but AMY's result for $\mu_{A}=0$ \cite{Arnold:2001ms}.

Therefore, the circular polarization asymmetry $A_{\pm\gamma}=\frac{\Gamma^{\rm odd}}{\Gamma^{\rm total}}\approx 0.03$ for $\omega/T=2$, $\alpha_{s}=0.2$, and $\mu_{A}/T=0.1$ which is about three times more than the strong coupling result $A_{\pm\gamma}\approx0.01$ that we found in \cite{Mamo:2013jda} using AdS/CFT correspondence.
 \begin{figure}[t]
 \centering
 \includegraphics[height=5cm]{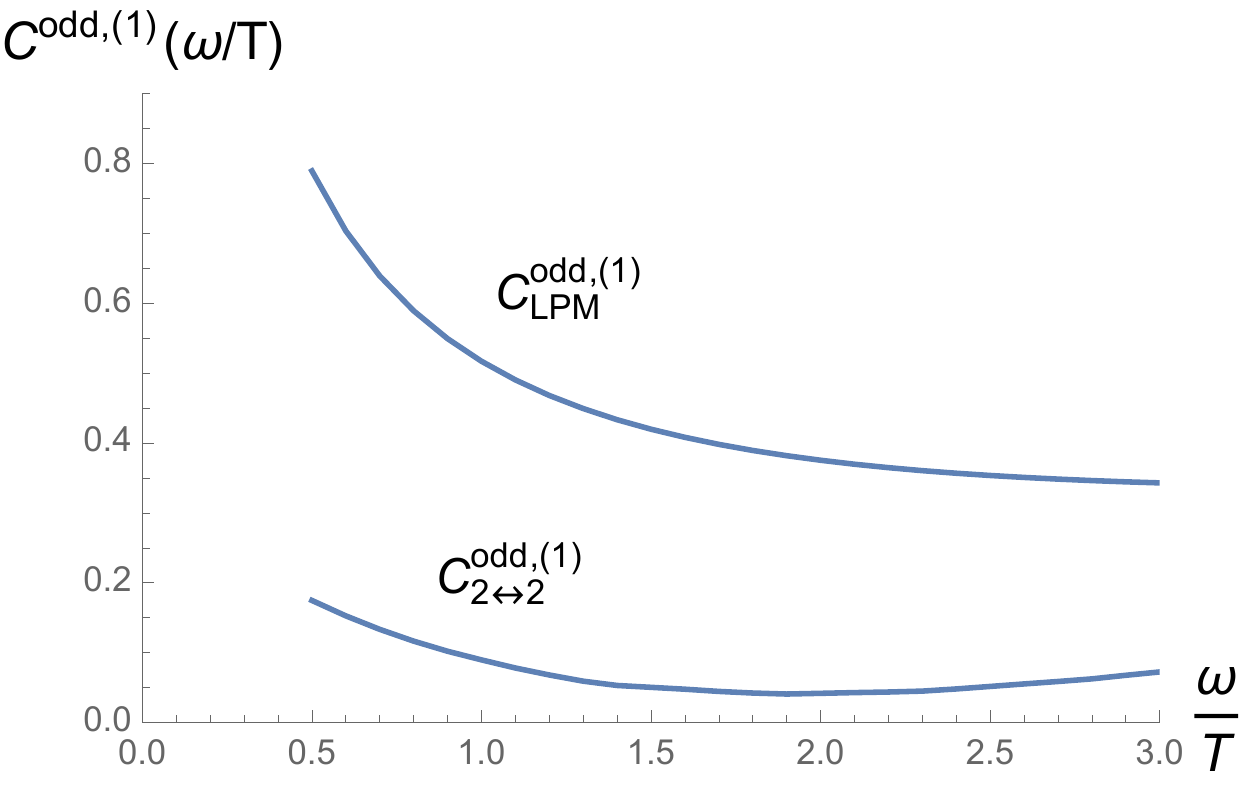}\caption{ Numerical results for $C^{\rm odd,(1)}_{2\leftrightarrow 2}(\omega/T)$, $C^{\rm odd,(1)}_{\rm LPM}(\omega/T)$ for $N_F=2$ QCD. \label{figFin}}
 \end{figure}

\vskip 1cm \centerline{\large \bf Acknowledgment} \vskip 0.5cm

We thank Meseret Demise, Jacopo Ghiglieri, J.-F. Paquet, Rob Pisarski, Matthew Siebert, and Derek Teaney for helpful discussions.

%\section*{Appendix 1 :}

%%%%%%%%%%%%%%%%%%%%%%%%%%%%%%%%%%%%%%%%%%%%%%%%%%%%%%%%%%%%%%%%%%%
\vfil

\end{document}